\documentclass[twocolappendix,trackchanges,twocolumn]{aastex7} 
\usepackage{amsmath}

\def\hi{H\,{\sc i}}
\def\vsig{$V/\sigma$}

\def\msol{M$_\odot$}
\def\Ngal{67 }
\def\Ngalbarba{32 }
\def\barba{Barba et al., in prep.}
\def\oiii{O\,{\sc iii}}

\newcommand{\R}[1]{\textbf{{\color{blue}{[DR:#1]}}}}

\begin{document}

\title{$V/\sigma$ Trends with Mass for Dwarf Galaxies from the Marvelous Massive Dwarfs Suite}

\author[orcid=0000-0002-8783-570X]{Dilys Ruan}
\affiliation{Department of Physics and Astronomy, Rutgers University, Piscataway, NJ 08854, USA}
\email[show]{druan@physics.rutgers.edu}  

\author[orcid=0000-0002-0372-3736]{Alyson M. Brooks} 
\affiliation{Department of Physics and Astronomy, Rutgers University, Piscataway, NJ 08854, USA}
\email{abrooks@physics.rutgers.edu}

\author[orcid=0009-0003-3401-3619]{Leonardo A. Barba}
\affiliation{Department of Astronomy, University of Massachusetts, Amherst, Amherst, MA 01003, USA}
\email{lbarba@umass.edu}

\author[orcid=0000-0002-4739-046X]{Mithi A. C. de los Reyes}
\affiliation{Department of Physics \& Astronomy, Amherst College, Amherst, MA 01002, USA}
\email{mdelosreyes@amherst.edu}

\author[orcid=0000-0001-7831-4892]{Akaxia Cruz}
\affiliation{Center for Computational Astrophysics, Flatiron Institute, New York, NY 10010, USA}
\affiliation{Department of Physics, Princeton University, Princeton, NJ 08544, USA}
\email{acruz@flatironinstitute.org}

\author[orcid=0000-0003-1509-9966]{Robel Geda}
\affiliation{Department of Astrophysical Sciences, Princeton University, Princeton, NJ 08544, USA}
\email{robel@princeton.edu}

\author[orcid=0000-0002-8040-6785]{Annika H. G. Peter}
\affiliation{CCAPP, The Ohio State University, Columbus, OH 43210, USA}
\affiliation{Department of Physics, The Ohio State University, Columbus, OH 43210, USA}
\affiliation{Department of Astronomy, The Ohio State University, Columbus, OH 43210, USA}
\email{peter.33@osu.edu}

\author[orcid=0000-0002-9642-7193]{Benjamin W. Keller}
\affiliation{Department of Physics and Materials Science, The University of Memphis, Memphis, TN 38152, USA}
\email{bkeller1@memphis.edu}  

\author[orcid=0000-0001-5510-2803]{Thomas Quinn}
\affiliation{Astronomy Department, University of Washington, Seattle, WA 98195, USA}
\email{trq@astro.washington.edu}  

\author[orcid=0000-0001-8745-0263]{James W. Wadsley}
\affiliation{Department of Physics and Astronomy, McMaster University, Hamilton, L8S 4M1, Canada}
\email{wadsley@mcmaster.ca}  

\begin{abstract}
    Galaxy formation scenarios can be interpreted through morphology and the level of rotational versus pressure support, quantified through the ratio of a galaxy's rotation speed to its velocity dispersion: $V/\sigma$. Observational studies of dwarf galaxies find that $V/\sigma$ does not strongly depend on environment, and may weakly depend on galaxy mass, which could shift our understanding of how dwarf galaxies form. We utilize the Marvelous Massive Dwarfs suite to examine whether $V/\sigma$ depends on mass in simulations, and understand how this varies for different baryonic components of the galaxy: \hi ~gas, young stars ($<$1 Gyr) and old stars ($>$1 Gyr). We use a simulation sample of 67 isolated dwarf galaxies with M$_\star=10^6-10^9$M$_\odot$ and produce line-of-sight kinematic maps for different viewing angles per galaxy. We find that $V/\sigma$ increases with mass, and that \hi ~gas and young stars are more rotation-supported ($V/\sigma\approx1-13$) while old stars are more dispersion-supported ($V/\sigma\approx0.2-5$). This result is consistent with the scenario where young stars are born from cold gas in the interstellar medium and undergo dynamical heating over time. We quantify the effects of spatial resolution on $V/\sigma$ and find that observations using old stars may underestimate the intrinsic $V/\sigma$. We find a correlation between $V/\sigma_\mathrm{HI,global}$ and \hi ~profile shape that is qualitatively similar to previous simulation results, but we find higher $V/\sigma_\mathrm{HI,global}$ compared to prior work which mostly found values $\lesssim2$ in this mass range. Our results motivate future work to examine $V/\sigma$ and dwarf galaxy formation with different kinematic tracers of the galaxy.
\end{abstract}


\section{Introduction} \label{intro}
We have yet to fully understand what influences angular momentum (`rotation') versus pressure (`dispersion') support in galaxies. The Milky Way exhibits structure that is kinematically and morphologically disky, while dwarf galaxies (M$_\star < 10^9$ M$_\odot$) seem to be more dispersion-dominated or irregular. One way to quantify the level of support is through kinematics with the ratio $V/\sigma$, where $V$ is the galaxy's rotation velocity and $\sigma$ is its velocity dispersion. The Milky Way exhibits $V/\sigma\sim 10$ when using measurements from either local \hi ~observations or measurements of thin disk stars, though other galaxies of comparable mass tend to have a stellar $V/\sigma\sim 5$ \citep{Pessa2023, McCluskey2025}. Meanwhile, dwarf galaxies exhibit $V/\sigma\lesssim 2$ when measured with stars \citep{2017Wheeler, 2023delosReyes}. These observations may suggest that there is a dichotomy of rotation-supported (high \vsig) versus dispersion-supported (low \vsig) galaxies, but they may also imply there should be a trend of increasing $V/\sigma$ with galaxy mass.  Either way, we have yet to fully understand how $V/\sigma$ varies as galaxy mass increases, and whether there is a sharp transition or a more gradual increase.

Meanwhile, to discern trends in $V/\sigma$ with galaxy mass, we must also disentangle the role of environment, particularly since many of the dwarfs with $V/\sigma$ data are within the gravitational influence of larger galaxies.  Earlier theoretical work predicted that disky dwarf galaxies are accreted and transformed into dwarf spheroidals \citep[e.g.,][]{2001MayerA,2001MayerB}. In this picture, ram pressure strips the infalling galaxy's gas, while repeated tidal perturbations alter its morphology, transforming dwarf irregulars (dIrrs) into gas-poor dwarf spheroidals (dSphs).  In this scenario, an evolution in $V/\sigma$ is expected, as rotation-dominated dIrrs are transformed into dSphs dominated by stellar dispersion ($\sigma_\star$) and $V/\sigma_\star < 1$ \citep[e.g.,][]{2009Klimentowski, 2010Mayer, 2012Lokas,2016Tomozeiu, Kazantzidis2017,2025Rathore}.

However, observations of stars in dwarf galaxies suggest that $V/\sigma_\star$ is not a function of environment, which has major implications for dwarf galaxy evolution. \citet{2017Wheeler} examined 40 dwarf galaxies from the Local Volume, while \citet{2023delosReyes} examined 30 more distant dwarf galaxies in the DIVE (``Dwarfs in Void Environments'') Survey. Neither study found clear trends between $V/\sigma_\star$ and distance to the nearest L$_\star$ galaxy (a proxy for environment). The lack of trend with environment suggests that dwarfs are born with lower $V/\sigma_\star$ than had been hypothesized, and that $V/\sigma_\star$ does not evolve significantly if dwarf galaxies indeed evolve from dIrrs to dSphs. 

Instead of depending on environment, observations find that $V/\sigma_\star$ might be a function of galaxy mass, though larger samples are needed to determine whether this trend is discrete or more gradual. By including the data from \citet{2017Wheeler}, \citet{2023delosReyes} found a hint of increasing $V/\sigma_\star$ over a mass range of M$_{\star}\sim10^4-10^9$ M$_\odot$.  Using a linear fit for $V/\sigma_\star$ versus $\log($M$_\star/M_\odot)$, they found a slope of $0.24 \pm 0.04$ and an intercept of $-0.98 \pm 0.35$. 

It is also likely that $V/\sigma$ varies based on the baryonic component used to measure $\sigma$, since stars are born from kinematically cold gas and expected to increase in dispersion with age \citep{1951Spitzer,1977Wielen, 2017Leaman}. The $\sigma$ measurements of all of the above studies in dwarf galaxies were dominated by older stellar populations.  To our knowledge, $V/\sigma$ has not been explicitly studied as a function of mass using the \hi ~gas of dwarf galaxies in observations.  However, this quantity can be determined with existing survey data by using the rotation speed (either V$_\mathrm{max}$ from resolved rotation curves or W$_{20}$/2 from unresolved \hi ~observations, which is the width at 20\% of the peak \hi ~emission divided by 2) and dividing by a global $\sigma_\mathrm{HI}$. Hence, observational constraints for $V/\sigma_\mathrm{HI}$ of dwarf galaxies can come from the kinematic analysis of galaxies from THINGS \citep[``The \hi ~Nearby Galaxy Survey'';][]{2008Walter}, LITTLE THINGS \citep[``Local Irregulars That Trace Luminosity Extremes, The \hi ~Nearby Galaxy Survey'';][]{2012Hunter,2015Oh} and VLA-ANGST \citep[``Very Large Array survey of Advanced Camera for Surveys Nearby Galaxy Survey Treasury galaxies'';][]{2012Ott}. In particular, global $\sigma_{\rm HI}$ values have been determined for galaxies in these surveys by \citet{2013StilpA} and \citet{Oman2019}. `Beam smearing' is a phenomenon where unresolved regions have artificially flatter velocity gradients, hence less rotation is inferred and this results in artificially higher velocity dispersions. High-resolution observations, like in the surveys listed above, generally minimize the effects of beam smearing \citep[as discussed in Section 6 of][]{2015Oh}.

Additionally, state-of-the-art HI surveys have begun taking data of dwarf galaxies, such as  Apertif \citep{2022vanCapellen,2022Adams}; FEASTS \citep[FAST Extended Atlas of Selected Targets Survey,][]{Wang2025}; MHONGOOSE \citep[MeerKAT \hi ~Observations of Nearby Galactic Objects: Observing Southern Emitters, ][]{2024deBlok}; MIGHTEE-\hi ~\citep[MeerKAT International GigaHertz Tiered Extragalactice Exploration survey - \hi,][]{2021Maddox}; and WALLABY \citep[Widefield ASKAP L-band Legacy All-sky Blind surveY,][]{2020Koribalski, 2020Spekkens, 2024Deg}.  These surveys are sensitive to galaxies with rotation speeds as low as $\sim30-40$~km~s$^{-1}$.
These surveys can make spatially resolved maps of diffuse \hi ~gas down to lower surface densities than ever before, and provide new opportunities to assess the unique physics of star formation in more low-mass galaxies. With more maps of rotation velocity and velocity dispersion, we can further our understanding of baryonic feedback and galaxy evolution for dwarf galaxies across a wider range of masses and morphologies.

Simulations provide a way to help us interpret velocity dispersions, both as a function of time and baryonic component (e.g., young stars versus old stars, or \hi ~gas).  Using the FIRE-2 simulations, \citet{McCluskey2024} found that $\sigma_\star$ depends on what component of the galaxy is being measured (specifically all stars or young stars), since stellar dispersions increase with age. The age-velocity dispersion relation of stars might be set by the initial dispersion of star-forming gas, which itself may change with time \citep{2007Kaufmann,2018ElBadry,Bird2013, Bird2021}.  Stars may also be dynamically heated over time due to mergers \citep{2012Helmi} or scattering with spiral arms or giant molecular clouds \citep{2015Kruijssen, 2017Leaman}, among other non-axisymmetric instabilities like bars \citep{2006Debattista}. The existing literature on dynamical heating mostly focuses on Milky Way-mass galaxies, but these mechanisms of dynamical heating could be mass-dependent. In the dwarf regime, we are still in the early stages of understanding whether these age-kinematic trends hold. For their sample of simulated dwarf galaxies, \citet{Benavides2025} compared the fraction of rotational to total kinetic energy ($\kappa_\mathrm{rot}$) between baryonic components and found that old stars are indeed more dispersion-supported, while cold gas and young stars exhibit more rotation support at similar levels. 

Simulations find that the level of rotation support increases with galaxy mass \citep[e.g.,][]{2018ElBadry,2018ElBadryHIprofiles,2019Dutton,Benavides2025,Celiz2025}. However, most simulations have had difficulty reproducing some observed aspects of dwarf galaxy morphology. For example, real dwarf galaxies exhibit an empirical size--mass relation \citep{VanderWel2011, Lelli2016, Carlsten2021,2025Asali}, but many simulations fail to make more compact galaxies, possibly due to strong feedback \citep{Jiang2019,Sales2022}. In addition, disk-dominated dwarf galaxies have been found in observations \citep[e.g.,][]{2012Ott,2015Oh,Lelli2016}, but stronger feedback may prevent the formation of disks in simulated dwarf galaxies.  Using the FIRE (``Feedback in Realistic Environments'') simulations, \citet{2018ElBadry,2018ElBadryHIprofiles} found that galaxies with stellar masses $<10^8$ M$_\odot$ do not form \hi ~gas disks and are more dispersion-supported with $V/\sigma_\mathrm{HI}\lesssim 2$. Similarly, \citet{Benavides2025} used FIREbox, a uniform volume simulation run with FIRE-2 physics, and did not find any disks (gas or stellar) for galaxies with M$_\star<10^9$ M$_\odot$. They highlighted that this transition mass is inconsistent with observations from LITTLE THINGS \citep{2015Oh}, in which dwarf galaxies with M$_\star < 10^8$ M$_\odot$ can exhibit $V/\sigma_\mathrm{HI}>3$. In another study using the Illustris TNG50 (``The Next Generation Illustris'') simulations, \citet{Celiz2025} found this transition from spheroid to disky galaxies at a slightly lower mass, between M$_\star\sim10^{8}-10^9$ M$_\odot$. However, they also cautioned that these simulated dwarf galaxies could be artificially more dispersion-supported due to limited resolution of their star formation physics, which preserves baryonic clumps around 1 kpc in size that prevent disk formation in the surrounding region. 

The high-resolution Marvelous Massive Dwarfs simulations  \citep[``Massive Dwarfs'',][]{Cruz2025} successfully reproduce characteristics of observed dwarf galaxies with stellar masses from $\sim 10^6 - 10^9$ M$_\odot$, and provide an opportunity to study $V/\sigma$ versus mass across baryonic components. \citet{Cruz2025} found that the Massive Dwarfs reproduce observed size--mass relations \citep{VanderWel2011, Lelli2016, Carlsten2021,2025Asali}, along with diversity of rotation curve shapes \citep{2015Oman}. In addition to this, the Massive Dwarfs exhibit disky morphologies down to at least M$_\star\sim10^8$ M$_\odot$ \citep{Keith2025, Geda2025}. The Massive Dwarfs are gas-rich, with gas-to-stellar mass fractions (1.4M$_\mathrm{HI}$/M$_\star)$ $\sim1-10$, and reproduce observed \hi ~velocities and sizes \citep{2025Ruan}. The simulated dwarfs also match the observed mass--metallicity relation and specific star formation rates of dwarfs, and they reasonably reproduce the column densities of low-ionization metals in the circumgalactic medium of low-redshift dwarfs \citep{Piacitelli2025}. Given this, the Massive Dwarfs allow us to study $V/\sigma$ for \hi ~gas and stars, across different dwarf galaxy morphologies.

In this work, we quantify the trend between $V/\sigma$ versus galaxy mass using different baryonic components (\hi ~gas, young stars, and old stars) for a sample of \Ngal galaxies from the Massive Dwarfs simulations. In Section \ref{methods}, we describe the simulations and methods used to derive global $V$ and $\sigma$ values. In Section \ref{globaltrends}, we present our results for $V/\sigma_\mathrm{global}$ using \hi ~gas and stars, both young and old.  In both cases we compare to observations. We also compare $V/\sigma_\mathrm{global}$ between baryonic tracers and compare our total sample versus subsamples of our galaxies that are disky or oblate. In Section \ref{sec:resolution}, we assess how $V/\sigma_\mathrm{global}$ changes with spatial resolution. In Section \ref{HIprofiles}, we examine the relation between \hi ~profile shape and $V/\sigma_\mathrm{HI,global}$. In Section \ref{discussion}, we discuss how the implementation of baryonic feedback in simulations could influence our results, along with comparisons to other simulations. We discuss what our results imply for the theory that field dwarfs transform into dwarf spheroidal satellites.  Finally, we summarize our work in Section \ref{conclusions}.

\section{Simulations \& Methods} \label{methods}
We utilize the high resolution ``zoom-in'' Marvelous Massive Dwarfs \citep{Cruz2025} simulations for this work. Zoom-ins re-simulate high-density regions of a halo and its surroundings from a larger-volume simulation. The region of the halo is simulated with higher-resolution particles and baryons, while the rest of the simulation box is kept at low resolution, dark matter-only. Large-scale structure is needed to deliver angular momentum to the zoom-in regions as described by tidal torque theory \citep{1969Peebles,1987Barnes}.  The Massive Dwarfs are higher-resolution runs of halos from the Romulus25 simulation \citep{2017Tremmel}. 40 zoom-in halos were selected based on two criteria: (i) being in the stellar mass range of M$_\star\sim10^{8-9}$ M$_\odot$ (where rotation curve diversity is maximized), and (ii) are at least 1.5 Mpc away from another galaxy with M$_\star >10^{10}$ M$_\odot$ at $z=0.1$. Another 15 zoom-in halos were selected to extend the mass range down to M$_\star\sim10^6$ M$_\odot$, with the same isolation criteria. The Massive Dwarfs have three times higher force resolution and 512 times higher mass resolution than Romulus25. Each Massive Dwarfs simulation has a volume of (25 Mpc)$^3$, and uses 87 pc for spline gravitational force resolution, 994  M$_\odot$ for initial star particle mass, 3300  M$_\odot$ for gas particle mass, and 18000  M$_\odot$ for dark matter particle mass. Massive Dwarfs and Romulus25 assume $\Lambda$CDM and parameters from Planck 2015 \citep[e.g., $h = 0.677$, $\Omega_m=0.307$, $\Omega_b h^2 =  0.02226$,][]{2016Planck}.

Massive Dwarfs were run using \textsc{ChaNGa} \citep{2015Menon}\footnote{\url{https://github.com/N-BodyShop/changa/}} for collisionless N-body and smoothed particle hydrodynamics (SPH) processes.  The hydrodynamics for \textsc{ChaNGa} have been imported from the tree+SPH code \textsc{GASOLINE2} \citep{2004Wadsley}.  \textsc{ChaNGa} is implemented in \textsc{Charm++} \citep{1993KaleKrishnan}, a parallel programming system, and handles dynamic load balancing to strongly scale with clustered data up to $\sim10^5$ cores. 

\textsc{ChaNGa} calculates gas cooling and non-equilibrium ion abundances by using collisional ionization rates \citep{1997Abel}, radiative recombination \citep{1981Black,1996Verner}, photo-ionization, free-free emission, and cooling from hydrogen and helium \citep{1992Cen}. A redshift-dependent, uniform UV background \citep{2012HaardtMadau} is included in the heating and cooling rates. We also simulate the non-equilibrium formation, shielding, cooling, and destruction of H$_2$, according to \citet{2012Christensen}. We use the metal-line cooling and metal distribution models from \citet{2010Shen}.

We adopt the stochastic star formation model from \citet{2006Stinson} and require the presence of H$_2$ based on \citet{2012Christensen}:
\begin{equation}
     p = \frac{m_\text{gas}}{m_\text{star}} \left( 1 - e^{-c_0^{*} X_{H_2} \Delta t/t_\text{form}} \right) \text{,}
\end{equation}
where $m_\text{gas}$ is the gas particle mass, $m_\text{star}$ is the initial star particle mass, $c_0^*$ is the star formation efficiency parameter (set to 0.1), $X_{H_2}$ is the H$_2$ mass fraction of the gas particle, $\Delta t$ is the time step for star formation in the simulation ($\sim$1 Myr), and $t_\text{form}$ is the local dynamical time. Star formation is limited to cold gas particles ($T<1000$ K), and a number density $n>0.1$ cm$^{-3}$ is required.  Star formation usually occurs in denser gas since we require the presence of H$_\mathrm{2}$, so the density for star formation is effectively $n > 100$ cm$^{-3}$. As shown in \citet{2014Christensen}, this star formation model reproduces the Kennicutt-Schmidt relation \citep{1959Schmidt,1998Kennicutt}. Each star particle in the Massive Dwarfs follows the initial mass function from \citet{2001Kroupa}.

Throughout each timestep, \textsc{ChaNGa} calculates the \hi ~mass fraction for each gas particle. The gas fraction is based on the particle's temperature and density, heating from the cosmic UV background radiation and young stars, H$_2$ self-shielding, and dust shielding in \hi ~and H$_2$. The total \hi ~mass of a galaxy (M$_\mathrm{HI}$) is calculated by summing over the product of the mass and \hi ~fraction for each gas particle. The total stellar mass (M$_\star$) and halo mass (M$_\mathrm{halo}$) are calculated by summing over particles within the virial radius.
\citet{2013Munshi} estimated that photometric observational methods yield stellar masses that are $\sim$60\% of the direct sum of simulated star particles within a halo, so we include this factor for a more direct comparison to observations. Throughout this study, any stellar mass will be 60\% of the simulated halo total. We also define the total baryonic mass as M$_\mathrm{bary}=$M$_\star + $1.4M$_\mathrm{HI}$, in which the 1.4 factor accounts for the mass of helium and metals in the gas phase \citep{1996Arnett}.

ChaNGa follows the return of energy, mass, and metals from stellar winds and Type\,Ia and Type\,II supernovae (SNe). Massive Dwarfs uses the superbubble feedback model from \citet{2014Keller} for Type\,II SNe, and deposits energy at a rate of $10^{51}$ erg per SN. Superbubble prevents numeric overcooling by placing feedback-heated particles into a temporary two-phase state, where each phase has a separate temperature and density. These two states correspond roughly to the cold swept-up ISM and the hot interior that has been heated by SNe and are kept in pressure equilibrium.  Mass flows from the cold phase to the hot via thermal evaporation \citep{1977Cowie} until the entire cold phase has been evaporated.  This evaporation process can continue by stochastically evaporating cool neighbors of hot, SN-heated gas particles.  Temperature gradients within resolved hot bubbles are smoothed by thermal conduction. Throughout our analysis, we exclude any particles in the temporary two-phase state.

We model the formation and evolution of black holes with a model designed to mimic the direct collapse of gas into massive black hole seeds \citep{Bellovary2011}. Black holes are formed via gas particles in regions of high density ($n > 1.5\times 10^4$ cm$^{-3}$), low temperature ($T<5\times10^3$ K), low metallicity ($Z<10^{-5}$), low molecular hydrogen fraction ($X_\mathrm{H_2} < 2\times10^{-3}$), and high Jeans mass ($M_J>10^5$ M$_\odot$).  These choices reproduce the conditions required to prevent fragmentation of gas. Black holes are seeded with a mass of $10^5$ \msol. The simulations include a modified Bondi accretion, following \citet{2017Tremmel} and motivated by \citet{2009Booth}, with a boost density threshold of 100 cm$^{-3}$. The accretion energy around a black hole is transferred to surrounding gas using the superbubble feedback model from \citet{2014Keller}. In order to understand the effects of black holes, we have a subset of zoom-ins which were run with the same fiducial physics but no black holes. In the 16 galaxies compared with and without black holes, the global velocity dispersions can vary up to 20\% but with no preferential shift. We conclude that the inclusion of black holes is not impacting our dispersion results.

The simulation data are analyzed with \textsc{Pynbody}\footnote{\url{https://pynbody.github.io/pynbody/}} \citep{2013Pontzen}, a python package for SPH + N-body simulations. We identify halos with \textsc{Amiga's Halo Finder} (AHF) \citep{2009Knollman}, and use the virial definition of 200 times that of the critical density at a given redshift.

\subsection{Galaxy Selection}

\begin{figure*}
    \centering
    \includegraphics[width=\textwidth]{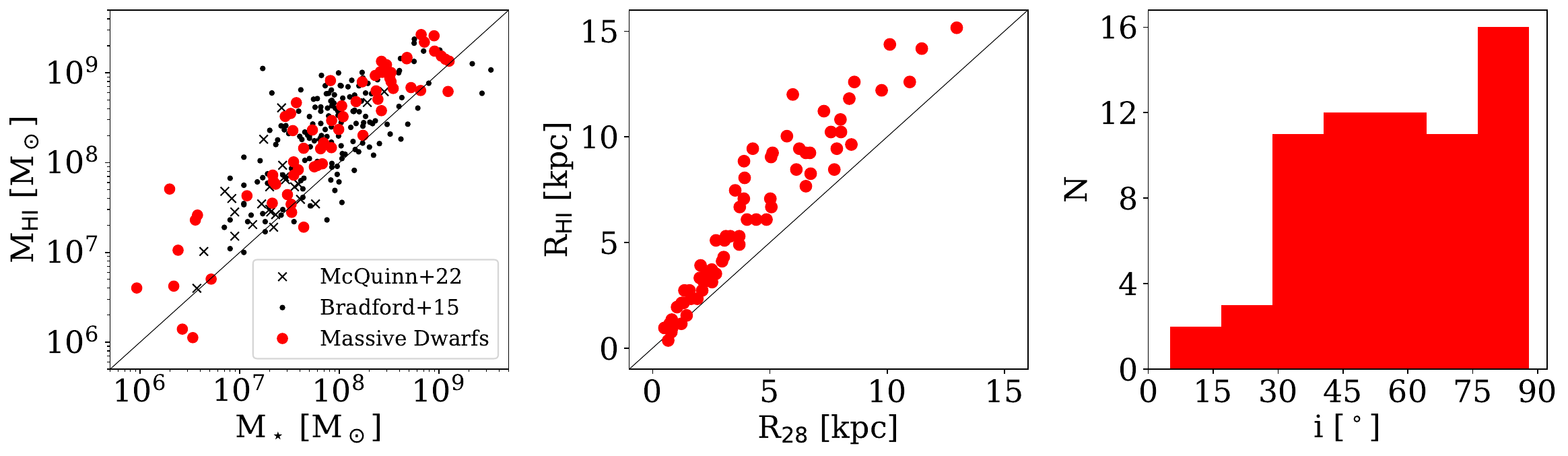}
    \caption{Left panel shows the \hi ~versus stellar mass relation for our sample of simulated galaxies (Massive Dwarfs, red markers) compared to observed galaxies from \citet{2022McQuinn} (black $\times$'s) and \citet{2015Bradford} (black points). Most of the simulated and observed galaxies have M$_{\rm HI} >$ M$_\star$ (above the one-one black line), i.e., these galaxies are gas-rich. The middle panel shows how the \hi ~versus stellar size limits imposed in our kinematic maps are comparable in range, but the \hi ~disk generally extends further than the stellar disk as R$_{\rm HI}>$ R$_{28}$. Right panel shows the distribution of inclination angles for our simulated dwarf galaxies. We randomly draw values of $\cos i$ from a uniform distribution, which results in higher inclinations for our sample on average. }
    \label{fig:sampleinfo}
\end{figure*}

    Our sample consists of \Ngal dwarf galaxies with stellar masses ranging from $\log($M$_\star/$M$_\odot)=5.97-9.09$, \hi ~masses from $\log($M$_\mathrm{HI}/$M$_\odot)=6.05-9.43$, and halo masses from $\log($M$_\mathrm{200}/$M$_\odot)=9.50-10.96$. This sample is selected to have M$_\mathrm{HI} > 10^6$ M$_\odot$, which corresponds to the detection limit of lower-mass galaxy surveys like SHIELD \citep[``Survey of \hi ~in Extremely Low-Mass Dwarfs'',][]{2011Cannon} and other local surveys \citep{2011Karachentsev}. In the left panel of Figure \ref{fig:sampleinfo}, we show the \hi ~versus stellar masses for our simulated sample (red markers). The black line shows a one-to-one relation and demonstrates how, on average, \hi ~mass is larger than stellar mass. Our masses are consistent with observed dwarf galaxies from \citet{2015Bradford} (black points) and \citet{2022McQuinn} (black $\times$'s). 

    As previously mentioned, there are 55 zoom-in runs, and we can analyze additional galaxies within the resolved Lagrangian region. We use a naming convention of ``r431-1'' to denote that the galaxy is from zoom-in run ``r431'', with a halo ID of 1 (which would be the most massive galaxy in the zoom region). 
    
    We aim to select isolated dwarfs whose kinematics are not strongly impacted by environment.  We select isolated field dwarfs based on the requirement that there are no other galaxies more massive than itself within a virial radius (R$_\mathrm{200}$).  We also removed two galaxies from our sample that undergo major mergers (mass ratio of 1:2) since $z\approx 0.1$. Our entire sample has not had any mergers with a mass ratio less than 1:10 since $z\approx 0.1$. 

    Our sample is selected by mass and exhibits a range of disky and non-disky morphologies. Following the definition described in Section 2.6.2 of \citet{Geda2025}, we classify these galaxies based on the normalized specific angular momentum of stars (${j}_{z,\star}/j_\mathrm{tot}$) when viewed face-on. We consider the galaxy `disky' if the median of ${j}_{z,\star}/j_\mathrm{tot}$ for all star particles is $> 0.5$. Based on this definition, 41 out of \Ngal of our simulated galaxies are disky.

    \subsection{Masking by Radius}

    We mask our data by radius to ensure satellites are not included in our kinematic maps. This radial cut also imposes a limit to study dwarf galaxy kinematics at similar sensitivities to observations.

    For maps of \hi ~gas, we limit our kinematic analysis out to the size at an \hi ~surface density limit of 1 \msol ~pc$^{-2}$ (R$_\mathrm{HI}$). In \citet{2025Ruan} we demonstrated that measuring out to a limit of 0.08 \msol ~pc$^{-2}$ is necessary for \hi ~to accurately trace the halo's maximum rotation speed and hence the turndown in the baryonic Tully-Fisher relation. However, we choose to measure out to 1 \msol ~pc$^{-2}$ in this study since it is the fiducial limit used in many radio surveys. 
    
    For kinematic maps of stars we consider data out to R$_\mathrm{28}$, which is the size at a B-band surface brightness of 28 mag arcsec$^{-2}$. This surface brightness limit was chosen as it measures out to $1$\% of the mean central surface brightness for our sample ($\sim 23$ mag arcsec $^{-2}$).  We calculate R$_\mathrm{28}$ with the profile class in \textsc{pynbody}. This method assumes stellar population models from \citet{2008Marigo} and \citet{2010Girardi}. 
    
    The middle panel of Figure \ref{fig:sampleinfo} compares the outer size limits used for the \hi ~maps versus the outer limit for stellar maps. On average, R$_\mathrm{HI}$ is greater than R$_\mathrm{28}$, as these values do not follow the one-to-one relation (black line). This is likely a consequence of the \hi ~disk extending farther out than the stellar disk, which is true in many observed galaxies \citep{1997Broeils,2008Begum,Lelli2016}. 

\subsection{Line-of-Sight Maps}
\label{LOSmaps}

First, we read in the particle data with \textsc{Pynbody} and center on the galaxy in position and velocity using the shrinking sphere method \citep{2003Power} for a region that is 1 kpc in size. We align each galaxy according to its midplane, which \textsc{Pynbody} determines from the total angular momentum of gas particles within 10 kpc from the halo's center. 

Then, we align the galaxy according to different inclination ($i$) and azimuthal ($\phi$) angles. For each galaxy, we generate a random value of $\cos i$ from a uniform distribution ranging from 0 to 1, convert $i$ to degrees, and use a random $\phi$. The right panel of Figure \ref{fig:sampleinfo} shows the distribution of inclination angles used throughout this work, resulting from our random draw in $\cos i$. In order to understand how much our kinematic quantities vary with viewing angle, we also generate maps with inclinations of $i=0^\circ,30^\circ,45^\circ,60^\circ,75^\circ,90^\circ$ in combination with azimuthal angles of $\phi=0^\circ,72^\circ,144^\circ,216^\circ,288^\circ$, resulting in a total of 31 viewing angles per galaxy. These maps are used for the errors in our global line-of-sight (LOS) kinematic quantities.

After centering and aligning the galaxy, we generate moment maps of column density (cm$^{-2}$); line-of-sight velocity (V$_{\rm LOS}$); and line-of-sight velocity dispersion ($\sigma_{\rm LOS}$). We bin particles in a uniform grid with spatial pixels (spaxels) that are 0.175 kpc in size. The spaxel length on each side is $\sim 2\epsilon_\mathrm{soft}$ (where $\epsilon_\mathrm{soft}$ is the spline gravitational force resolution,  $0.087$ kpc). For column density maps, we sum over particles for the total mass per spaxel, divide by the spaxel area, and then convert to cm$^{-2}$. For the first and second moment maps, we perform a mass-weighted average for V$_{\rm LOS}$ and $\sigma_{\rm LOS}$ per spaxel. We construct maps for each baryonic component: \hi ~(neutral hydrogen with $T<10^4$ K), young stars ($<$ 1 Gyr in age), and old stars ($>$ 1 Gyr in age).

\begin{figure*}
    \centering
     \raisebox{0.12\textwidth}{\includegraphics[width=0.24\textwidth]{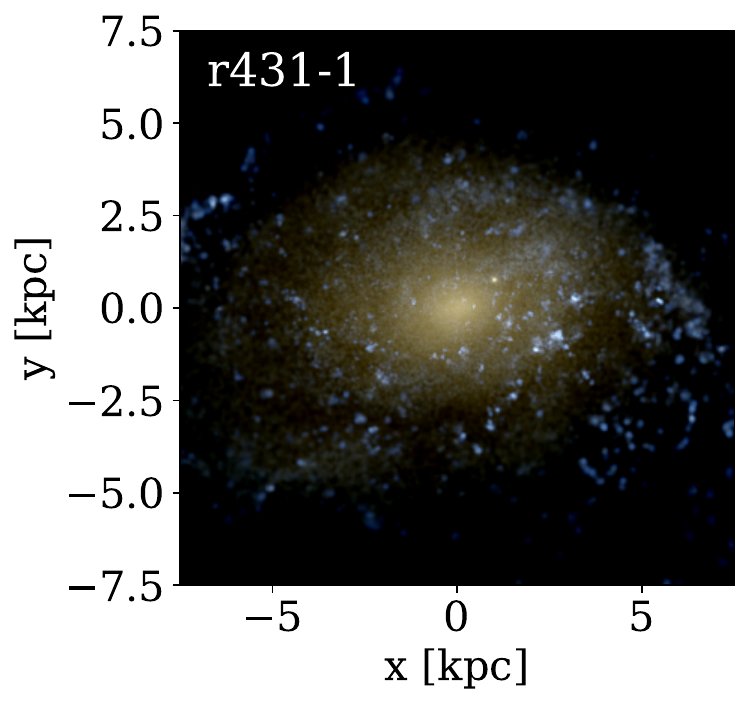}} 
    \includegraphics[width=0.75\textwidth]{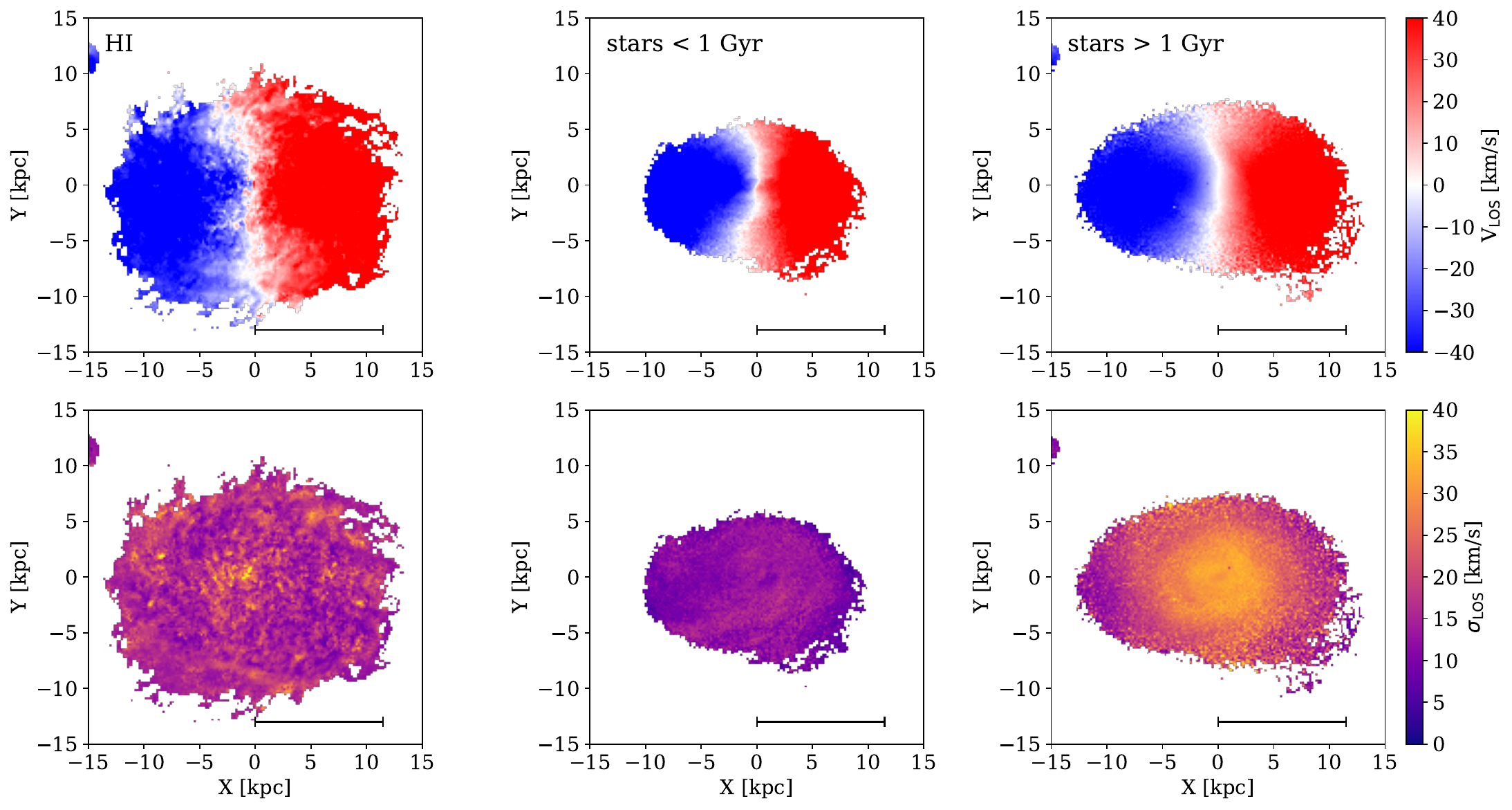}
    \raisebox{0.12\textwidth}{\includegraphics[width=0.24\textwidth]{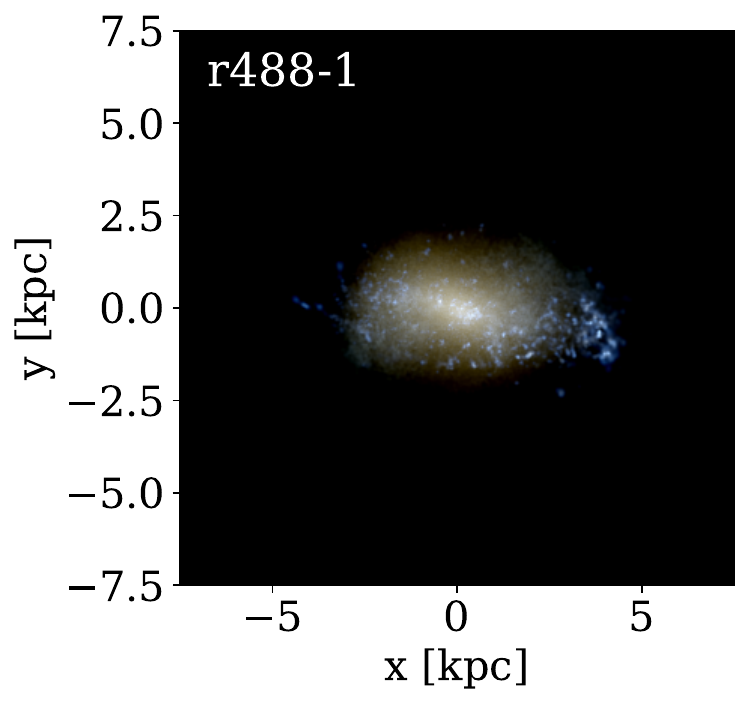}}
    \includegraphics[width=0.75\textwidth]{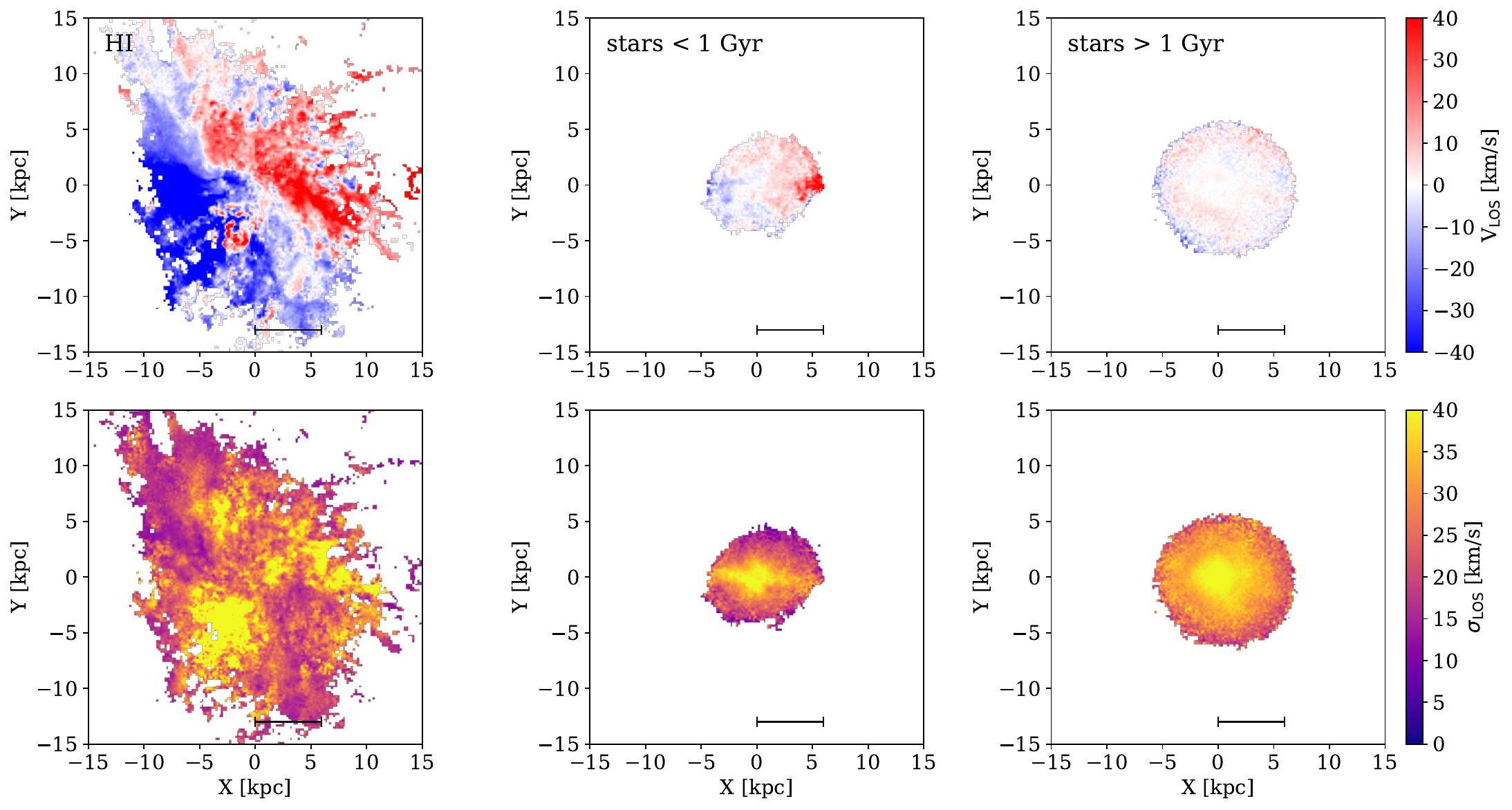}
    \caption{Left-most panels show mock UVI images of each galaxy oriented at an inclination angle of 45$^\circ$. The right-most twelve panels show maps of LOS velocity (blue-red color map) and LOS velocity dispersion (purple-yellow color map) at an inclination of 45$^\circ$. The black line on the bottom of each panel shows how we mask by radius, specifically R$_\mathrm{HI}$ for \hi ~gas and R$_\mathrm{28}$ for stars. The top six panels show maps for r431-1, a disky galaxy, based on using \hi ~(left), young stars (middle), and old stars (right) as tracers. The bottom six panels show maps for r488-1, an irregular galaxy. In both galaxies, the \hi ~and old stars are more extended than young stars. For a disky galaxy like r431-1, with $\log$(M$_\star/$M$_\odot$)=8.95, gas and young stars exhibit rotation and old stars exhibit higher dispersions. r431-1 has a satellite, as seen in the upper-left, which is masked out after the cut by radius. For an irregular galaxy like r488-1, with $\log$(M$_\star/$M$_\odot$)=9.02, the \hi ~and stars can exhibit different kinematics. r488-1 has rotating, disturbed \hi ~gas, while old stars and young stars do not exhibit as much rotation and have smoother dispersion fields. To track these galaxies throughout the analysis, we will denote results for r431-1 with a black `$\times$' and r488-1 with a black `+' in subsequent plots. \label{fig:maps}}
\end{figure*}

We show examples of these velocity and dispersion maps in Figure \ref{fig:maps}. The upper two rows show maps for r431-1, an extended disky galaxy, at an inclination angle of $i=45^\circ$. In terms of the spatial extent for each component, old stars and \hi ~gas extend further out than young stars.\footnote{It is common for dwarfs to be dominated by young stars in their central regions and older stars in their outskirts; see \citet{Riggs2024} and discussion therein.} As demonstrated in the $\sigma_\mathrm{LOS}$ row for r431-1 in Figure \ref{fig:maps}, older stars tend to have higher velocity dispersions than young stars or \hi ~gas.  Older stars can become more dispersion-dominated due to dynamical heating (see Appendix \ref{sub:disp_v_time}). In the lower two rows of Figure \ref{fig:maps}, we show maps of LOS velocity and dispersion for r488-1, which is an irregular galaxy. The \hi ~gas of r488-1 exhibits rotation and disturbed regions with high velocity dispersion. The old stars and young stars of r488-1 do not exhibit rotation as strongly, and the corresponding dispersion fields are more smooth compared to the \hi.

\subsection{Global V and $\sigma$}
\label{global}
We calculate a global value of V$_\mathrm{LOS}$ map using the relation
\begin{equation}
    V_\mathrm{global} = \frac{1}{2}(V_\mathrm{max} - V_\mathrm{min})\text{,}
    \label{eq:Vrot}
\end{equation}
where $V_\mathrm{max}$ and $V_\mathrm{min}$ are the maximum and minimum speeds in the first moment map (LOS velocity). This relation from \citep{2009Law} is used in other studies with integral field unit (IFU) data,
since flux-weighted or mass-weighted speeds can bias the rotation velocity in brighter central regions for non-elliptical galaxies \citep{2022Fraser}. If rotation speeds are weighted heavier in more central regions, this could result in an underestimate of the rotation speed.  

Typically with IFU data, a global $\sigma$ is flux-weighted according to the equation outlined by \citet{2003Cappellari} and \citet{2005Binney}. In order to have a consistent method between \hi ~gas and stars for our simulated galaxies, we instead use a mass-weighted global value. Previous simulation work on pressure versus rotation support in galaxies also use mass-weighted global values \citep[e.g.,][]{2018ElBadryHIprofiles,McCluskey2025,Benavides2025}. Since \hi ~emission can be assumed to be optically thin, a mass-weighted average is equivalent to intensity-weighted moment maps. We calculate a global LOS velocity dispersion ($\sigma_\mathrm{global}$) with the median of the mass-weighted second moment maps.

\section{$V/\sigma$ trends with mass} \label{globaltrends}
In this section we study how the amount of rotation versus pressure support depends on the mass of dwarf galaxies. First, we compare the $V/\sigma_\mathrm{global}$ values between our simulated galaxies and observational constraints for \hi ~gas and stars separately. Then, we compare $V/\sigma_\mathrm{global}$ between baryonic tracers. Finally, we examine whether morphologically disky galaxies are correlated with higher values of $V/\sigma_\mathrm{global}$. The data for LOS rotation speed, LOS velocity dispersion, and $V/\sigma_\mathrm{global}$ for the simulated galaxies have been made available on a GitHub repository \footnote{\url{https://github.com/dilysruan/v-sigma-maps}}.

\subsection{HI Gas}
\label{HI}
\begin{figure*}
    \centering
    \includegraphics[width=0.49\textwidth]{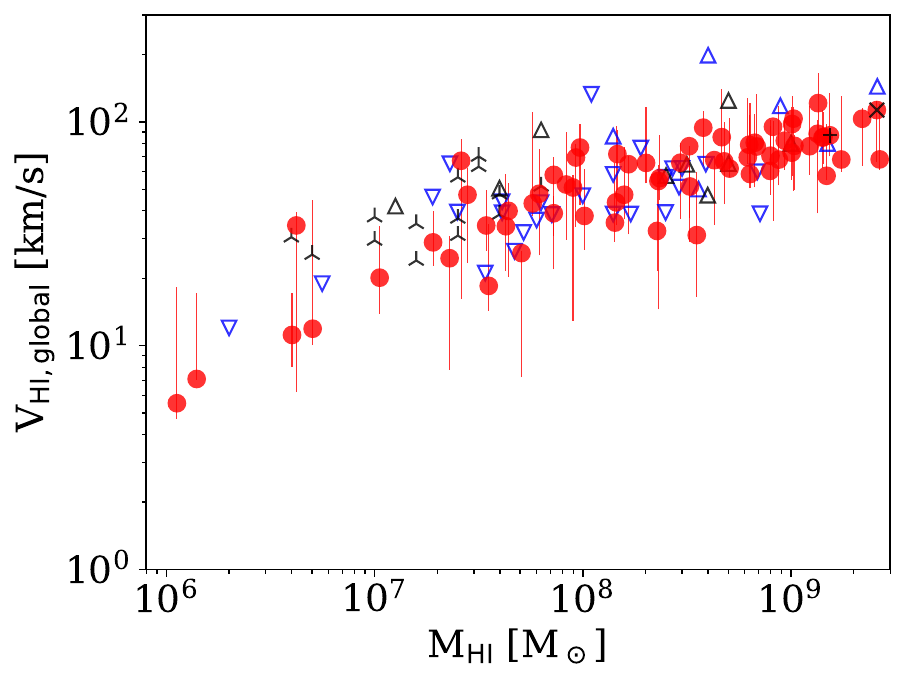}
    \includegraphics[width=0.49\textwidth]{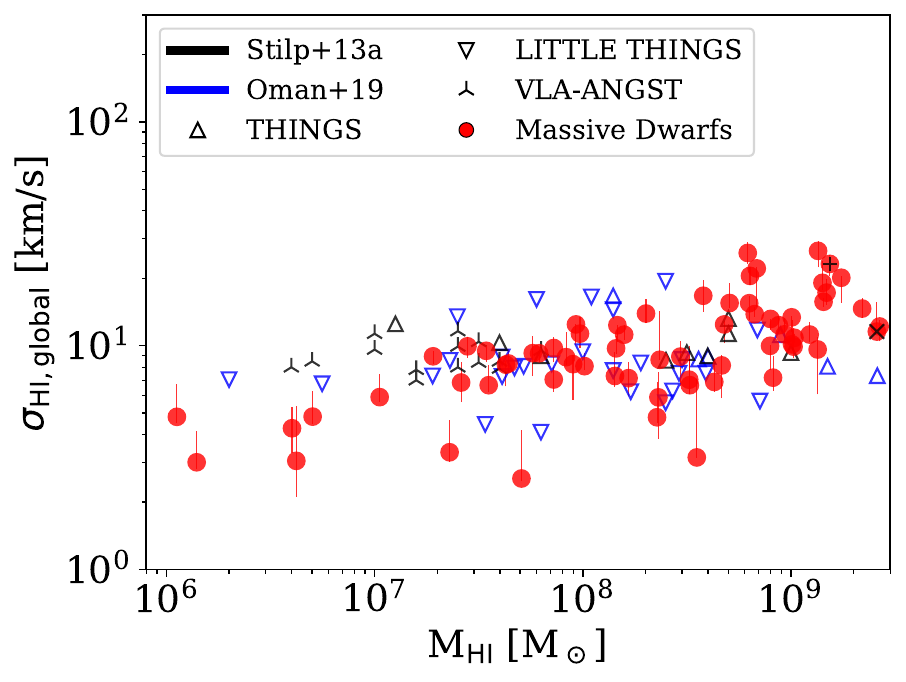}
    \includegraphics[width=0.65\textwidth]{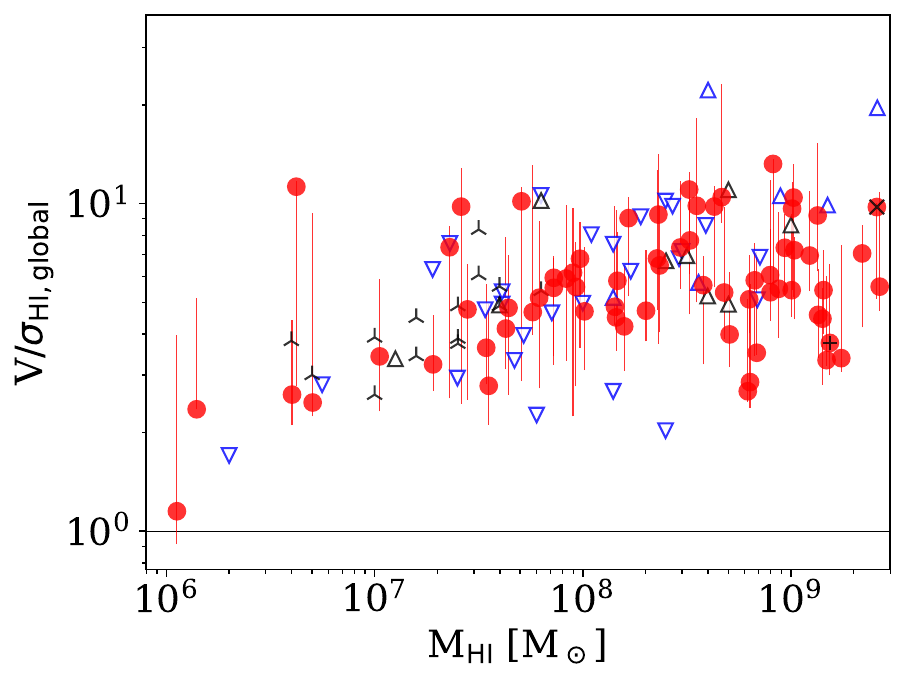}
    \caption{In each panel, the global \hi ~kinematic value of our simulated galaxies from the Massive Dwarfs suite (red markers) are measured at a random inclination. The vertical error bars extend from the minimum to maximum value over all 31 viewing angles. The left panel shows the \hi ~LOS velocity as a function of \hi ~mass.  We compare the simulation results to observed rotation speeds: V$_\mathrm{max}$ from THINGS and LITTLE THINGS used in \citet{Oman2019} (blue markers) and W$_\mathrm{20}/2$ from VLA-ANGST and THINGS used in \citet{2013StilpA} (black markers). There is overlap between \citet{2013StilpA} and \citet{Oman2019} for five galaxies: DDO 53, DDO 154, NGC 2366, DDO 154, NGC 7793, and IC 2574, and we show constraints from both works. In each panel, the black `$\times$' shows the kinematic quantity for r431-1 (a disky galaxy in our simulated sample) and the black `$+$' shows the result for r488-1 (an irregular galaxy), both of which are shown in Figure \ref{fig:maps}. The right panel compares the global LOS velocity dispersion versus M$_\mathrm{HI}$. Generally our V$_\mathrm{HI,global}$ and $\sigma_\mathrm{HI,global}$ values are consistent with observations. Finally, in the bottom panel we show the global LOS $V/\sigma$ versus M$_\mathrm{HI}$. Our simulated sample has consistent $V/\sigma_\mathrm{HI,global}$ values as observed galaxies in the same mass range, in which the \hi ~gas disks are mostly rotation-supported ($V/\sigma > 1$, shown with the horizontal black line). }
    \label{fig:HI_Vsigma}
\end{figure*}

 In Figure \ref{fig:HI_Vsigma}, we compare the global values of rotation velocity (left panel), velocity dispersion (right panel), and V$/\sigma_{\rm HI, global}$ (bottom panel) as a function of \hi ~mass. Our simulated galaxies are shown in the red markers, and the vertical error bars in each panel extend to the minimum and maximum of each kinematic quantity over all 31 viewing angles previously mentioned ($i=0^\circ,30^\circ,45^\circ,60^\circ,75^\circ,90^\circ$ and $\phi=0^\circ,72^\circ,144^\circ,216^\circ,288^\circ$). The red markers show values measured at a random inclination angle (from Figure \ref{fig:sampleinfo}), drawn from a uniform distribution in $\cos i$. We compare to global values for real galaxies measured by \citet{Oman2019} (blue markers) and \citet{2013StilpA} (black markers), with marker shapes representing different surveys: THINGS \citep[][]{2008Walter}, LITTLE THINGS \citep[][]{2012Hunter}, and VLA-ANGST \citep[][]{2012Ott}.  
 
 The left panel of Figure \ref{fig:HI_Vsigma} shows the well-established trend where rotation velocity increases with \hi ~mass. Our rotation speeds are determined from Equation \ref{eq:Vrot} using the maximum and minimum values in the LOS velocity map. For the observational data, we compare to V$_\mathrm{max}$ values from Table A2, Column 10 in \citet{Oman2019} and W$_\mathrm{20}/2$ values from Table 1, Column 10 in \citet{2013StilpA}. These rotation speeds are corrected for inclination, in which the observed linewidth is related to the true value through the approximation W$_\mathrm{20,obs} = $W$_\mathrm{20,intrinsic}\sin i$. For a more direct comparison, we inclination-correct our simulation values (from Equation \ref{eq:Vrot}) by dividing by $\sin i$ in the regime where $i\geq 40^\circ$, since the approximation is not valid at lower inclinations. Overall we find consistent results for the LOS rotation speed from our simulated maps and observations, with some observed galaxies exhibiting higher rotation speeds ($> 80$ km s$^{-1}$ at M$_\star \gtrsim10^8$ M$_\odot$) than our simulated sample. 

The right panel of Figure \ref{fig:HI_Vsigma} shows the global \hi ~velocity dispersions versus M$_\mathrm{HI}$. As previously mentioned, the simulated $\sigma_\mathrm{HI,global}$ values are calculated through the median of the mass-weighted second moment maps. The THINGS and LITTLE THINGS data were also determined in \citet{Oman2019} using the median of the second moment maps (see their Figure 3), while the VLA-ANGST data were determined from the intensity-weighted average \citep{2013StilpA}.

Discrepancies in the LOS velocity dispersion in simulations versus observations may arise due to how thermal and kinematic contributions are treated. The simulated \hi ~velocity dispersions are lower than some observed galaxies at the lower-mass end (M$_\mathrm{HI} \lesssim 10^7$ M$_\odot$), and higher than some observed galaxies at the higher-mass end (M$_\mathrm{HI} \gtrsim 5\times10^8$ M$_\odot$), with a difference of up to $\sim 5$ km s$^{-1}$ in both cases. Throughout this work, we only use kinematic contributions to the velocity dispersion. \citet{Oman2019} removed thermal contributions and only use kinematic contributions. \citet{2013StilpA} stated (in their Section 4.3) that the velocity dispersion of the central Gaussian ($\sigma_\mathrm{central}$, used in this comparison) can be physically interpreted as a result of random turbulent motions rather than thermal contributions, since the thermal energy of ISM gas in the temperature range $4000-12,000$ K radiates away on too short of timescales ($\sim10^3$ yr). Although kinematic contributions likely dominate the velocity dispersion, it is not as easy to separate kinematic and thermal dispersion components in observations. Therefore, the kinematic dispersions from our simulated galaxies may not exactly match observed dispersions, especially at the low-mass end, since it is more difficult to distinguish thermal and kinematic contributions. 

The global \hi ~rotation speed and velocity dispersion generally increase with \hi ~mass, with rotation speed having a stronger dependence on galaxy mass. In the bottom panel of Figure \ref{fig:HI_Vsigma}, we additionally show that V$/\sigma_\mathrm{HI,global}$ increases with M$_\mathrm{HI}$. Our simulated values are consistent with the range of observed V$/\sigma_\mathrm{HI,global}$. We find that the \hi ~gas is rotation-dominated for most of the simulated sample, as V$/\sigma_\mathrm{HI,global}$ is greater than 1 (shown with the horizontal black line) and ranges from V$/\sigma_\mathrm{HI,global}\approx 1-13$.

We discuss more on how the simulation physics may impact velocity dispersion in Section \ref{feedback}. Spatial resolution can also influence our result. As discussed in Section \ref{sec:resolution}, lower spatial resolution results in a bias towards lower rotation speeds and higher dispersions, which could be the case when comparing observations to simulated maps with 0.175-kpc resolution.

\subsection{Young and Old Stars}
\label{young_old_stars}

\begin{figure*}
    \centering
    \includegraphics[width=0.49\textwidth]{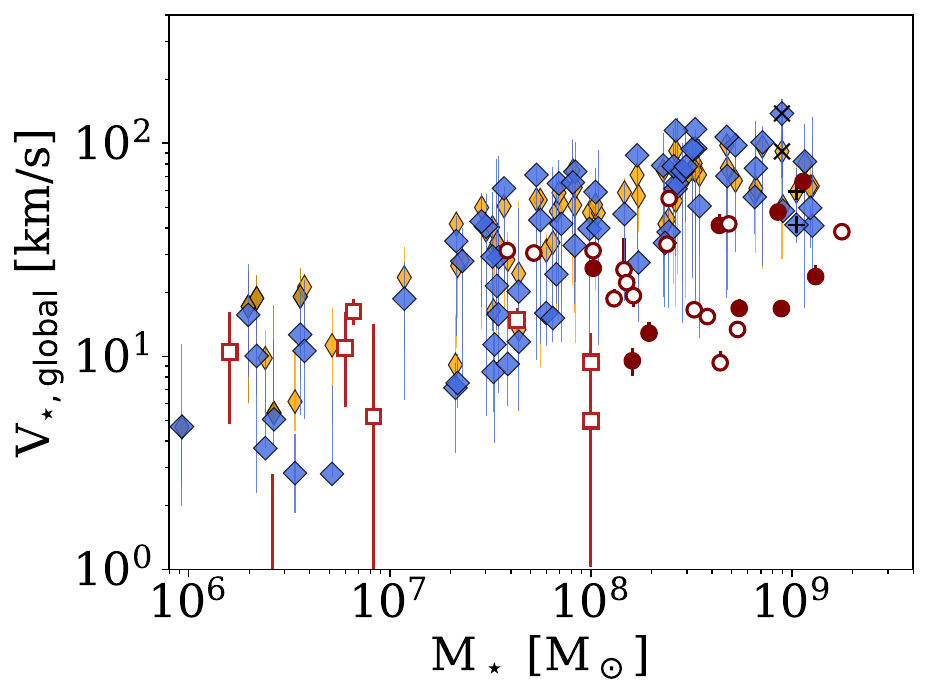}
    \includegraphics[width=0.49\textwidth]{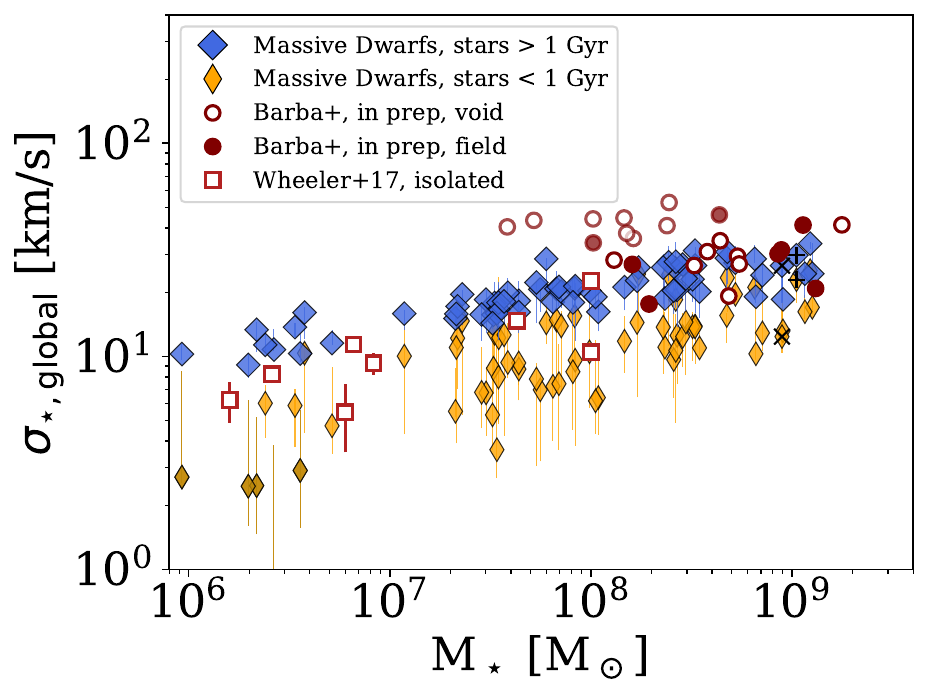}
    \includegraphics[width=0.65\textwidth]{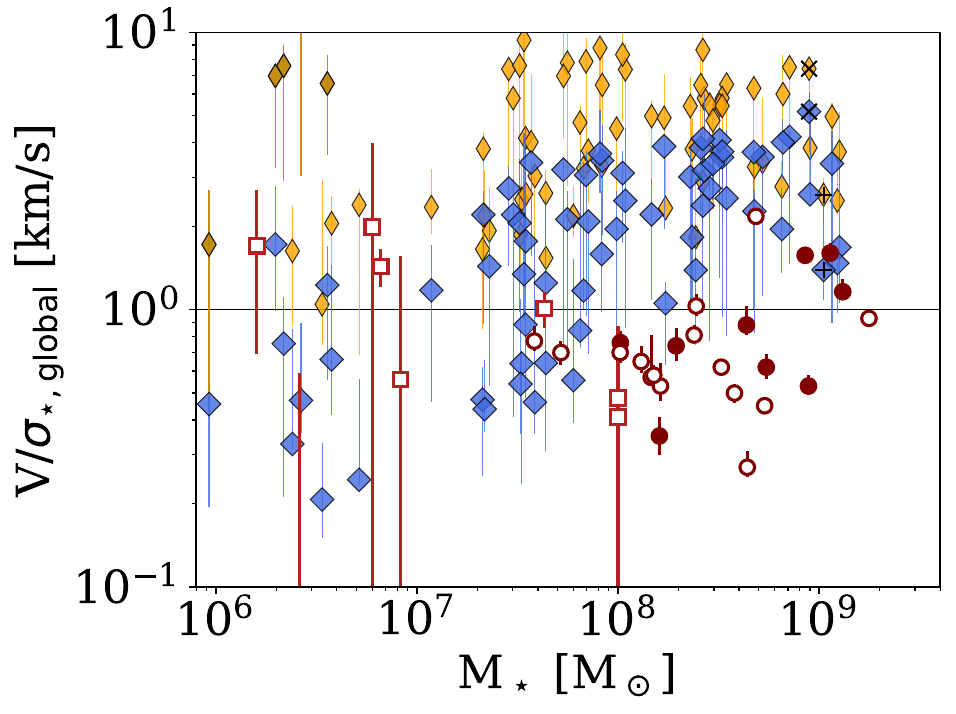}
    \caption{Similar to Figure \ref{fig:HI_Vsigma} but for the old stars (ages $>$ 1 Gyr, shown in blue diamonds) and young stars (ages $< 1 $ Gyr, shown in orange) in our simulated dwarf galaxies. Galaxies which have young stellar maps with $\leq10$ particles per spaxel are shown in dark yellow. We compare to observations from \citet{2017Wheeler} (open red squares) and \barba ~(maroon circles; open for void galaxies, filled for field galaxies). Each kinematic quantity from simulated galaxies is taken at a randomly-drawn viewing angle, and the vertical error bars extend from the minimum and maximum value across all 31 viewing angles. The left panel demonstrates how the global LOS velocity in simulations is generally higher than in observations, likely as a result of better spatial resolution in the simulation data with our idealized moment maps. We examine the effect of spatial binning in Section \ref{sec:resolution}. In the right panel, we find that old stars have higher LOS velocity dispersions than the young stars of the same galaxy. The dispersion range between old and young stars from simulated galaxies is generally consistent with observed values. However, 10 out of \Ngalbarba of the observed galaxies (with lower opacity markers) have global stellar velocity dispersions which are higher (by a factor of up to 2.3) than what we find in simulations for old stars at a given stellar mass. We discuss potential sources of this discrepancy in Section \ref{stellar_dispersions}. The bottom panel demonstrates how higher dispersions for the older stars results in lower global $V/\sigma$ values than for young stars. The horizontal black line shows $V/\sigma_{\star,\mathrm{global}} = 1$, below which most of the observed galaxies are dispersion-dominated, whereas only some of the simulated galaxies are dispersion-dominated at M$_\star < 10^8$ M$_\odot$. It is possible that this discrepancy in $V/\sigma$ is a result of different methodology rather than a physical result of galaxy support (Section \ref{sec:resolution}). }
    \label{fig:stars_Vsigma}
\end{figure*}

Using a similar approach, we compare $V/\sigma$ for stars in our simulated galaxies versus observations. We compare to observational constraints from \citet{2017Wheeler} and \barba, which is a follow-up paper to \citet{2023delosReyes}. We select isolated galaxies from \citet{2017Wheeler} with M$_\star \geq 10^6$ M$_\odot$, which provides more constraints on the lower-mass end. \barba ~has a sample of isolated dwarf galaxies in field and void environments with M$_\star = 10^7-10^9$ M$_\odot$, which overlaps well with our simulated sample. 

Different methodologies are adopted by \citet{2017Wheeler} and \barba ~in order to get LOS rotation speeds and LOS velocity dispersions. \citet{2017Wheeler} determine the rotation speed and velocity dispersion of their galaxies through Bayesian analysis with rotating and non-rotating models to fit their individual stellar spectra. \barba ~determine stellar velocities and dispersions by fitting their galaxy spectra with stellar templates and constructing IFU maps, then determining global values from the maps. Absorption features in the stellar continuum are dominated by older stars. Therefore, a more direct comparison between simulated and observed $V/\sigma$ should use old stars, and we use an age cut of $> 1$ Gyr. In this section, we also show results for young stars ($<$ 1 Gyr) to compare with the kinematics of old stars.

The left panel of Figure \ref{fig:stars_Vsigma} shows the stellar LOS rotation speed as a function of M$_\star$. Similar to Figure \ref{fig:HI_Vsigma}, the markers are values measured at a random inclination angles, drawn from a uniform distribution in $\cos i$, while the error bars show the range of values over all 31 viewing angles. Our simulated values for old stars are shown with blue diamonds, and young stars are shown with orange diamonds. Data from \barba ~are shown with maroon circles (open markers for galaxies in voids, and filled markers for field galaxies), and data from \citet{2017Wheeler} are the red open squares. The observed rotation speeds are not corrected for inclination, and therefore we also do not correct for inclination in Figure \ref{fig:stars_Vsigma}.

The right panel of Figure \ref{fig:stars_Vsigma} shows the stellar LOS velocity dispersion as a function of M$_\star$. For our simulated galaxies, old stars exhibit systematically higher dispersions that are, on average, more than $ 2\times$ that of young stars. The simulated values are mostly consistent with the observed stellar velocity dispersions  within the range between young and old stars. Ten of the \Ngalbarba real galaxies (marked with lower-opacity markers) exhibit dispersions that are higher by at least 10 km s$^{-1}$ (up to a factor of 2.3) than our simulated values at a given mass. 

The bottom panel of Figure \ref{fig:stars_Vsigma} shows $V/\sigma_{\star,\mathrm{global}}$ with stellar mass. Following the upper panels, old stars have higher velocity dispersions and similar rotation speeds as young stars, and therefore this results in lower $V/\sigma_{\star,\mathrm{global}}$ values. The old stars in our simulated dwarf galaxies exhibit $V/\sigma_{\star,\mathrm{global}}\approx 0.2-5$. As shown with the horizontal black line for $V/\sigma=1$, more of the old stars in our simulated dwarf galaxies become dispersion-dominated below M$_\star\sim10^8$ M$_\odot$. Observed galaxies seem to be more dispersion-supported compared to our simulated $V/\sigma$ values for old stars. 19 out of the \Ngalbarba real galaxies have $V/\sigma_{\star,\mathrm{global}} < 1$, whereas 16 out of the \Ngal simulated galaxies have $V/\sigma_{\star,\mathrm{global}} < 1$.

Spatial resolution and sample selection may explain differences between the simulations and observations. In Section \ref{sec:resolution}, we explore how lower (uniform) spatial binning and Voronoi binning may influence V$_{\star,\mathrm{global}}$ and $\sigma_{\star,\mathrm{global}}$. We find that lower uniform spatial resolution means worse sampling of the velocity field, and hence lower rotation speeds and higher velocity dispersions. Voronoi binning results in larger spatial bins in the outer regions of the galaxy, and hence lower rotation speeds and lower dispersions. It is likely that differences in stellar $V/\sigma_{\star,\mathrm{global}}$ between simulations and observations stem from methodology. However, higher stellar dispersions for some galaxies in \barba ~cannot be fully explained by Voronoi binning. In Section \ref{stellar_dispersions}, we discuss what aspects might differ in galaxy populations between the simulations and observations to result in comparatively higher stellar velocity dispersions for some of the observed galaxies in the right panel of Figure \ref{fig:stars_Vsigma}. 

\subsection{Comparing $V/\sigma$ between baryonic tracers}

\begin{figure}
    \centering
    \hspace{-5mm}{\includegraphics[width = \linewidth]{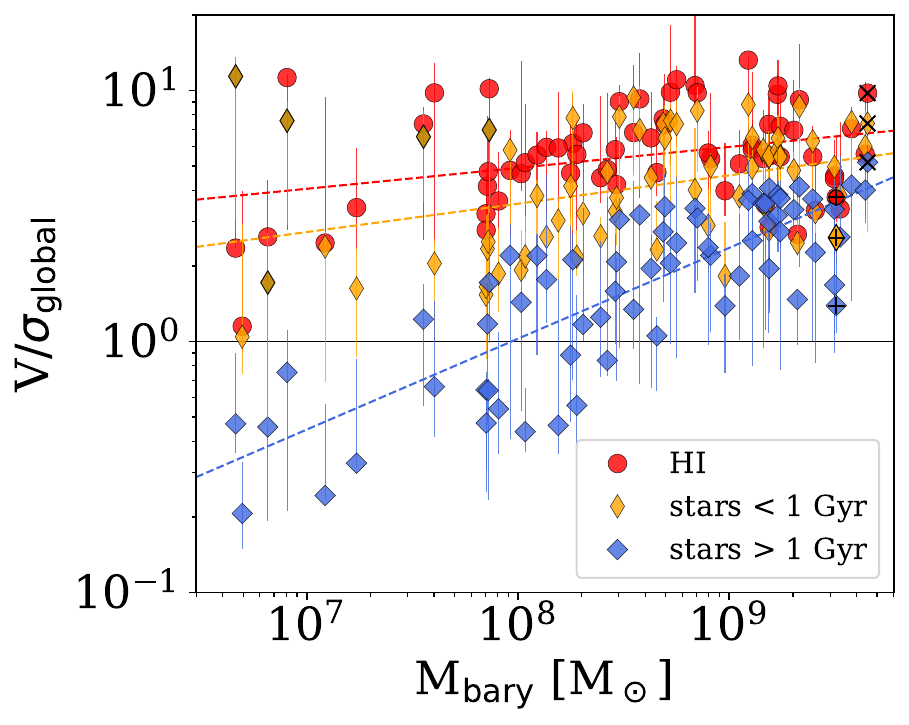}}
    {\includegraphics[width = 0.9\linewidth]{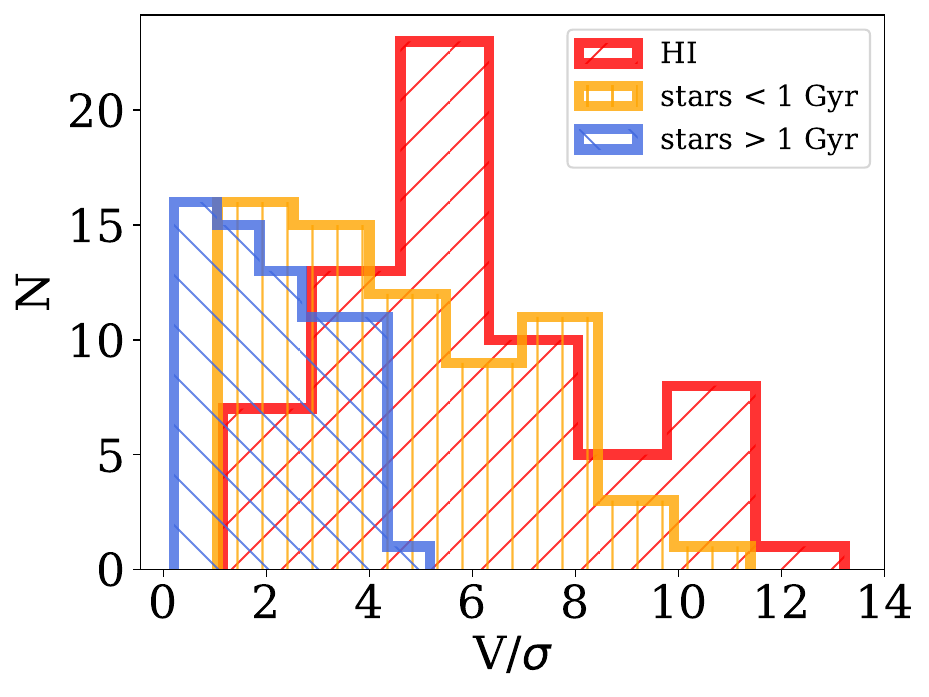}}
    \caption{Top panel shows $V/\sigma$ versus total baryonic mass for each kinematic tracer (\hi ~- red circles, old stars - blue diamonds, young stars - orange diamonds). We define M$_\mathrm{bary}=1.4$M$_\mathrm{HI} + $M$_\star$. Similar to Figures \ref{fig:HI_Vsigma} and \ref{fig:stars_Vsigma}, we show the $V/\sigma$ at a randomly-drawn viewing angle, and the vertical error bars extend from the minimum and maximum value across all 31 viewing angles. The dark yellow diamonds denote galaxies which have young stellar maps with $\leq10$ particles per spaxel. The dotted lines show power-law fits to the data, with the parameters in Table \ref{tab:vsigma_fit}. The bottom panel shows histograms of $V/\sigma$. The \hi ~and young stars in our galaxies exhibit $V/\sigma \approx 1-13$, whereas old stars exhibit $V/\sigma\approx 0.2-5$. These results are consistent with our picture of galaxy formation: \hi ~is more rotation-supported (higher $V/\sigma$) and cools to form young stars, which then become more dispersion-supported over time (lower $V/\sigma$). Our result demonstrates how the global $V/\sigma$ depends on which baryonic component is being studied for a given galaxy, and that the level of rotation support scales with galaxy mass.}
    \label{fig:Vsigma_all}
\end{figure}

In Figure \ref{fig:Vsigma_all}, we compare the global values of $V/\sigma$ for our simulated sample of galaxies and their respective baryonic components (cold \hi ~gas, young stars, and old stars). The top panel of Figure \ref{fig:Vsigma_all} shows $V/\sigma$ as a function of total baryonic mass (M$_\mathrm{bary} =$ 1.4M$_\mathrm{HI} + $ M$_\star$). The dotted lines represent a power-law fit of the form $y=Ax^b$, in which $x=$M$_\mathrm{bary}$ and $y=V/\sigma$. The fit parameters, root-mean-squared (RMS) error, and scatter in $V/\sigma_\mathrm{global}$ are listed in Table \ref{tab:vsigma_fit}.

The bottom panel of Figure \ref{fig:Vsigma_all} shows the distribution of $V/\sigma$ for each component. \hi ~and young stars are more rotation-supported (with $V/\sigma\approx 1-13$), while old stars are more dispersion-supported (with $V/\sigma\approx0.2-5$).  The young stars and \hi ~gas may exhibit similar levels of rotation support due to how young stars are born from cold gas in the ISM. Over time, we expect stars to undergo dynamical heating, which will increase their velocity dispersions over cosmic time, and this is consistent with our trend for older stars. As a proof-of-concept, in Appendix \ref{sub:disp_v_time} we demonstrate dynamical heating for a few disky galaxies from our sample.

\begin{table}
    \centering
    {\begin{tabular}{|c|c|c|c|c|}
    \hline
        Component & $b$ & $A$ & RMS & STD\\
        \hline
        \hi &  $0.083 \pm 0.03$ & $1.06 \pm 0.26$ & 2.59 & 2.58\\
        \hline 
        stars $<$ 1 Gyr & $0.11 \pm 0.04$ & $0.44 \pm 0.31$ & 2.35 & 2.34 \\ 
        \hline 
        stars $>$ 1 Gyr & $0.36 \pm 0.03$ & $0.001 \pm 0.28$ & 0.83 & 1.25 \\
        \hline
    \end{tabular}}
    
    \caption{Best-fit parameters of our data from Figure \ref{fig:Vsigma_all} to the power-law $y=Ax^b$, in which $x=M_\mathrm{bary}$ and $y=V/\sigma$. The RMS error (RMS) compares the power-law fit to simulation data, and the last column shows scatter in $V/\sigma$ through the standard deviation (STD). Errors for each fit parameter ($b,A$) are determined using the diagonals of the covariance matrix, with a scaling factor of $\chi^2/N$, where $N$ is the number of degrees of freedom.}
    \label{tab:vsigma_fit}
\end{table}

\begin{figure*}
    \centering
    \hspace{-5mm}{\includegraphics[width=0.435\linewidth]{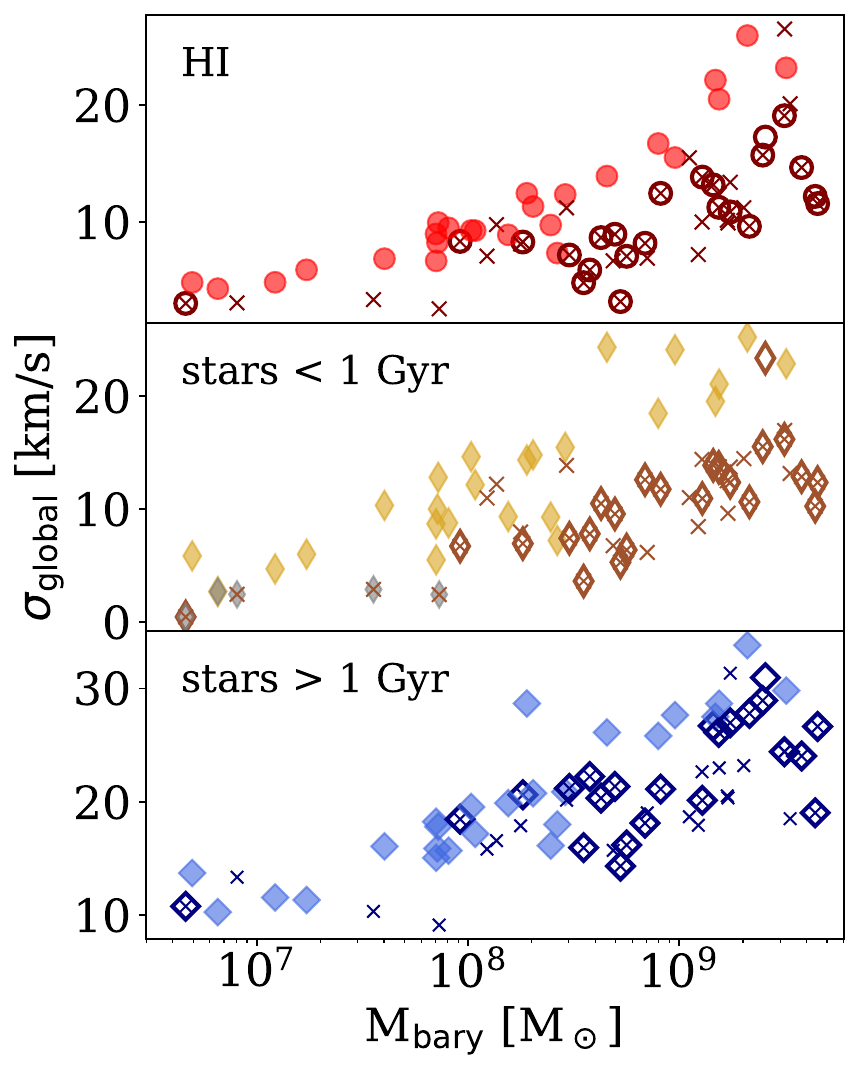}}
    \raisebox{1mm}{\includegraphics[width=0.455\linewidth]{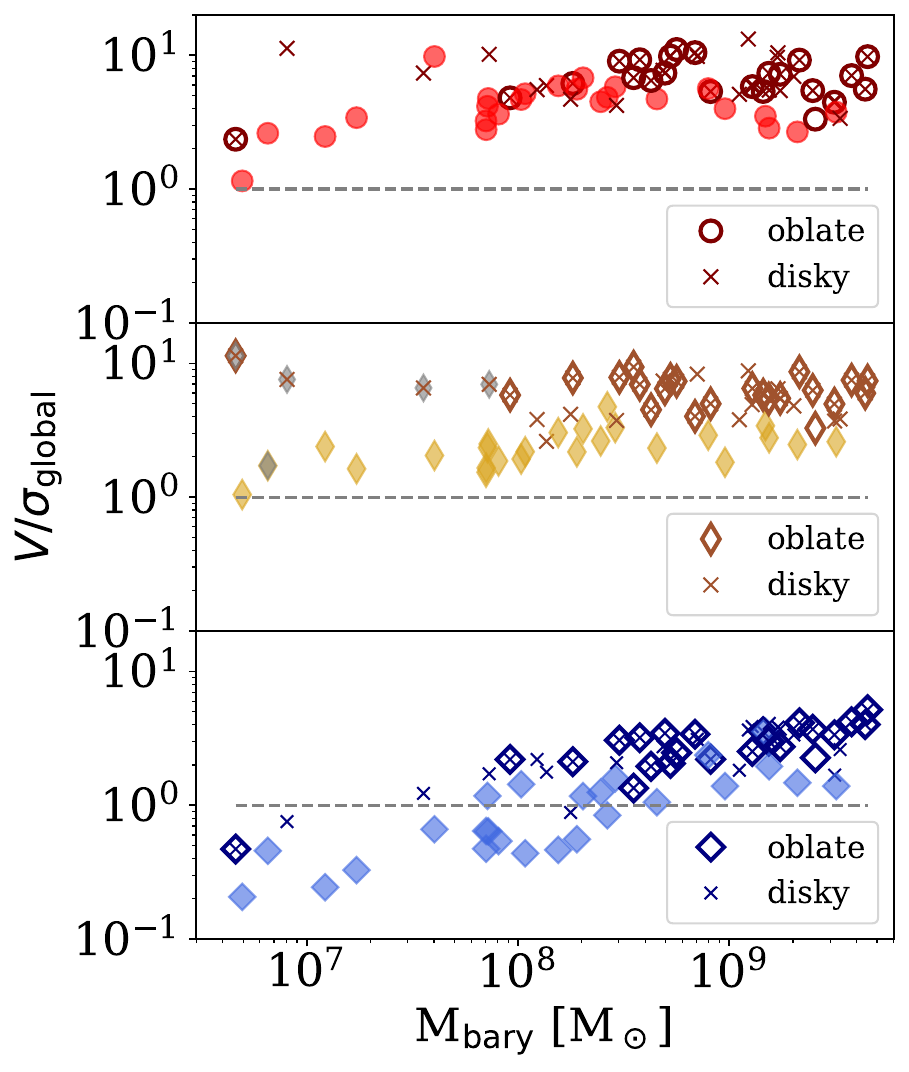}}
    \caption{We assess whether higher $V/\sigma_\mathrm{global}$ is correlated with disky or oblate galaxies. Each row shows a different baryonic component: \hi ~gas (top row, red), young stars (middle row, yellow), and old stars (bottom row, blue). A galaxy is considered disky based on the specific angular momentum vector of its stars (median ${j}_{z,\star}/j_\mathrm{tot} > 0.5$). Galaxies are oblate if the triaxiality of its stars is $T<1/3$.  The grey diamonds denote galaxies which have young stellar maps with $\leq10$ particles per spaxel. In the left column, we compare $\sigma_\mathrm{global}$ versus baryonic mass for disky and oblate galaxies. Both disky (`$\times$' markers) and oblate (empty markers) galaxies exhibit systematically lower $\sigma_\mathrm{global}$ at a given baryonic mass, and this is especially apparent in the dispersions of \hi ~gas and young stars. In the right panel, we compare $V/\sigma_\mathrm{global}$ for disky galaxies and oblate galaxies. The grey dashed line shows $V/\sigma_\mathrm{global} = 1$. Disky/oblate galaxies indeed exhibit higher $V/\sigma_\mathrm{global}$. Despite this correlation, a galaxy with higher $V/\sigma_\mathrm{global}$ and strong rotation does not exclusively mean that it is disky/oblate (see r488-1 in Figure \ref{fig:maps}). }
    \label{fig:disky}
\end{figure*}

By comparing $V/\sigma$ across baryonic components of the galaxy, we gain a more complete picture for how to interpret kinematic signatures with respect to galaxy formation. dSph galaxies with low $V/\sigma$ could be formed from material which was already dispersion-dominated, or they could be a result of dIrr galaxies evolving over time. \citet{2017Wheeler} and \citet{2023delosReyes} measured $V/\sigma$ with absorption features (dominated by old stars) and found a lack of trend with environment, which could suggest that dSphs start as dispersion-dominated objects. However, our result demonstrates that a given galaxy may exhibit different $V/\sigma$ values depending on which type of stellar population or gas is being observed. In summary, the baryonic component and galaxy mass should be considered when measuring the level of rotation versus pressure support for a galaxy.  We discuss the implications further in Section \ref{sec:scenarios}.

\subsection{$V/\sigma$ and disky morphologies}

We consider whether strongly-rotating galaxies with higher $V/\sigma_\mathrm{global}$ match morphologically disky galaxies. Throughout this work, we have considered a galaxy to be disky based on the specific angular momentum of its stars, where the median ${j}_{z,\star}/j_\mathrm{tot} > 0.5$ \citep[following][]{Geda2025}. Disks can also be defined through axis ratios of the galaxy. \citet{Keith2025} found that the triaxiality of stars and triaxiality of the dark matter halo are both correlated with whether a galaxy is considered disky by visual inspection. Triaxiality is defined as the following:
\begin{equation}
    T = \frac{1-Q^2}{1-S^2}\text{,}    
\end{equation}
where $Q = B/A$ and $S=C/A$, and $A$, $B$, and $C$ are the semi-major, semi-intermediate, and semi-minor axes of an ellipsoid, respectively. Disky galaxies tend to be `oblate', with $0 < T < 1/3$. Using \textsc{pynbody}, we align the galaxy face-on and calculate the axis ratios $B/A$ and $C/A$ for star particles as a function of radius. Then, we determine each axis ratio at two times the effective radius (2R$_\mathrm{eff}$) and compute $T$.

In the left column of Figure \ref{fig:disky}, we compare $\sigma_\mathrm{global}$ versus baryonic mass for disky galaxies and oblate galaxies versus the entire simulated sample. Each row shows the result for a given baryonic component, where the top row is \hi, middle row is young stars, and bottom row is old stars. The diskiness definitions based on specific angular momentum (`$\times$' markers) and triaxiality (empty markers) produce similar results, but do not exactly match. Disky and oblate galaxies exhibit lower velocity dispersions than other galaxies at a given baryonic mass, and this is especially apparent in \hi ~gas and young stars. 

\begin{figure*}[ht!]
    \centering
   \includegraphics[width=\textwidth]{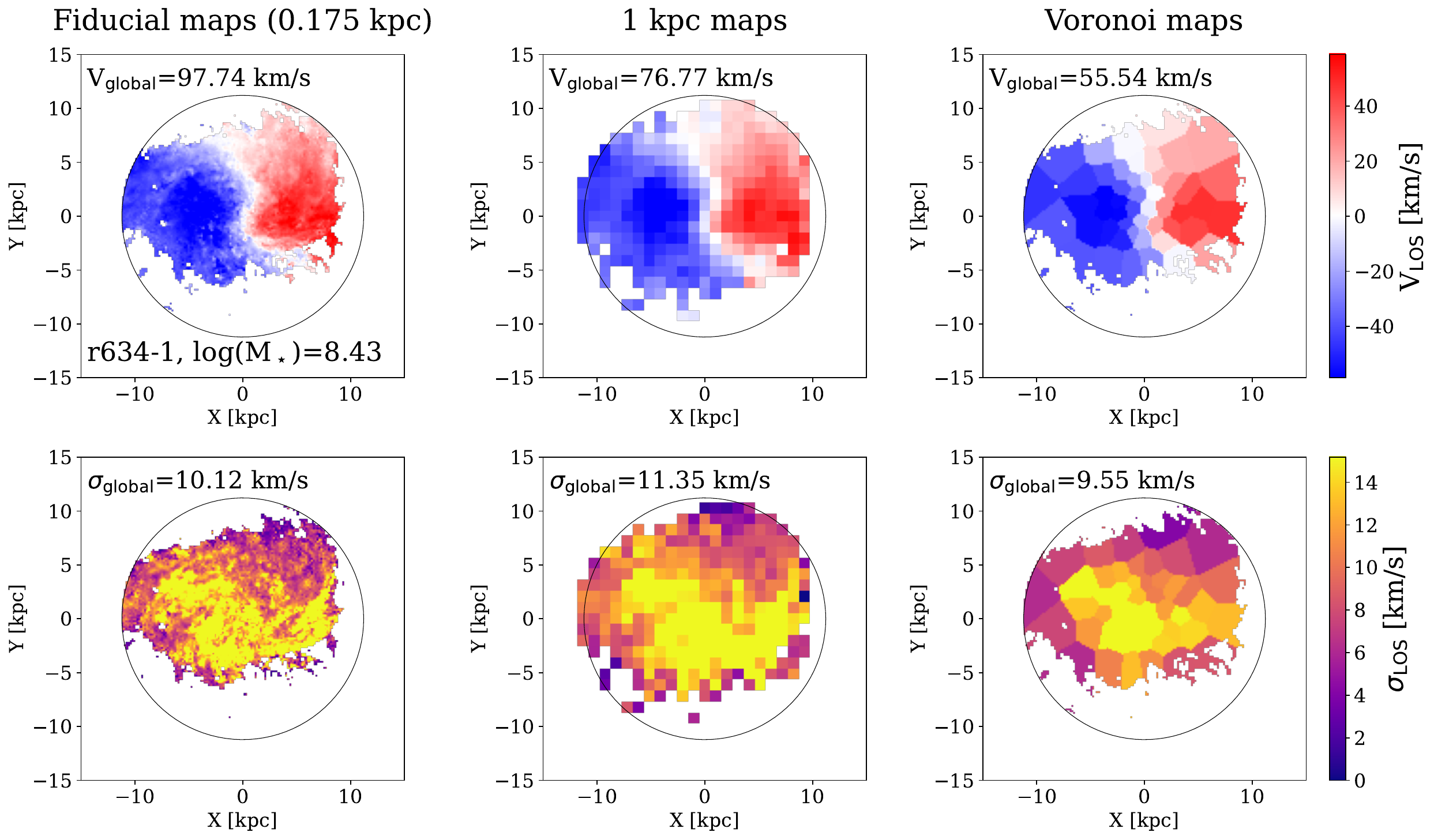}
\caption{Kinematic maps of \hi ~gas for r634-1 (a disky galaxy at $i=52^\circ$). We compare the global values of LOS velocity and LOS velocity dispersion with high spatial resolution (0.175 kpc spaxels; left column), low spatial resolution (1 kpc spaxels; middle column), and Voronoi binning (variable bin sizes; right column). Top row: $V_\mathrm{global}$ can be underestimated as larger spaxels will `average out' the LOS velocities per spaxel to a lower value. Voronoi binning typically results in larger spatial bins in the outer regions of the galaxy, and thus the global rotation velocity is lower than that of the 1-kpc map. Bottom row: similar to beam smearing, worse spatial resolution results in an artificially higher global velocity dispersion, as shown in the 1-kpc map. However, with Voronoi binning, spatial bins are larger in the outer regions of the galaxy which have lower dispersion, and therefore this decreases the global value.}
    \label{fig: binsize}
\end{figure*}

In the right column of Figure \ref{fig:disky}, we compare $V/\sigma_\mathrm{global}$ for disky galaxies and oblate galaxies. We find that indeed, galaxies which have disks according to their stellar specific angular momentum and oblateness exhibit higher $V/\sigma_\mathrm{global}$. 

Despite this correlation between disky/oblate galaxies and $V/\sigma_\mathrm{global}$, it seems that identifying a disk through $V/\sigma_\mathrm{global}$ alone is not as straightforward. $V/\sigma_\mathrm{global} > 1$ suggests that a galaxy is rotation-dominated, though this does not perfectly correspond to a disky galaxy given the definitions we have explored here. For example, r488-1 (shown in Figure \ref{fig:maps}) exhibits $V/\sigma_\mathrm{global}=1.39$ for old stars and $V/\sigma_\mathrm{global}=3.75$ for \hi. However, r488-1 is non-disky in terms of $j_{z,\star}/j_\mathrm{tot}$ and $T$. A galaxy may exhibit strong rotation, but its rotation is not necesarily confined to the shape of a disk.

\section{Systematic Effects from Spatial Binning}\label{sec:resolution}

In Figure \ref{fig:stars_Vsigma}, we found systematically lower $V/\sigma_{\star,\mathrm{global}}$ values between observed galaxies with IFU data compared to the simulations. In order to understand how spatial binning factors into our result, we use lower spatial resolution and Voronoi binning and compare the global kinematic values. Variable bin sizes can be used with IFU data of stellar kinematics to reach a certain level of signal-to-noise \citep[e.g.,][Barba et al., in prep.]{2023delosReyes}. We compare the kinematic maps when our uniform square pixel size is changed from 0.175 kpc on each side (used throughout this work), to 1 kpc on each side, and to varying spatial bins with the \verb|vorbin| algorithm \citep{2003Cappellari}. For Voronoi maps, we use mass maps (normalized by their mean value) as the input signal, take the square root of this signal assuming Poisson statistics for the input noise, and set a target S/N (same value for a given baryonic component) in order to achieve $\geq 10$ bins per galaxy. We demonstrate this Voronoi binning in the GitHub repository\footnote{\url{https://github.com/dilysruan/v-sigma-maps}} accompanying this work. 

\begin{figure*}
    \centering
    \includegraphics[width=\textwidth]{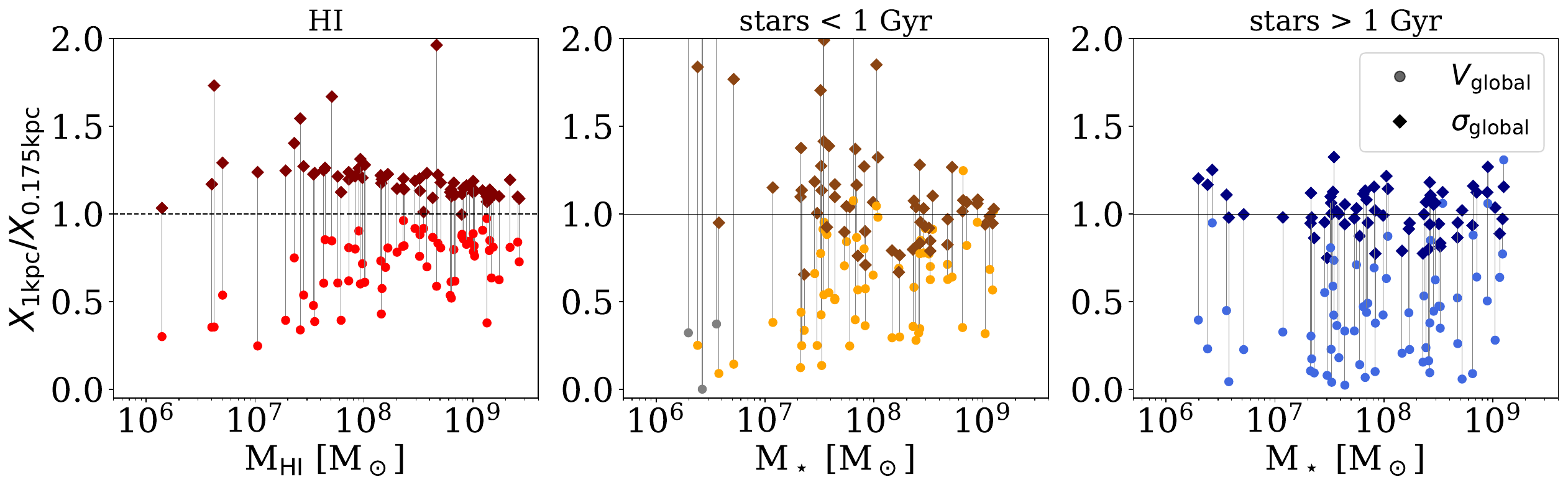}
    \includegraphics[width=\textwidth]{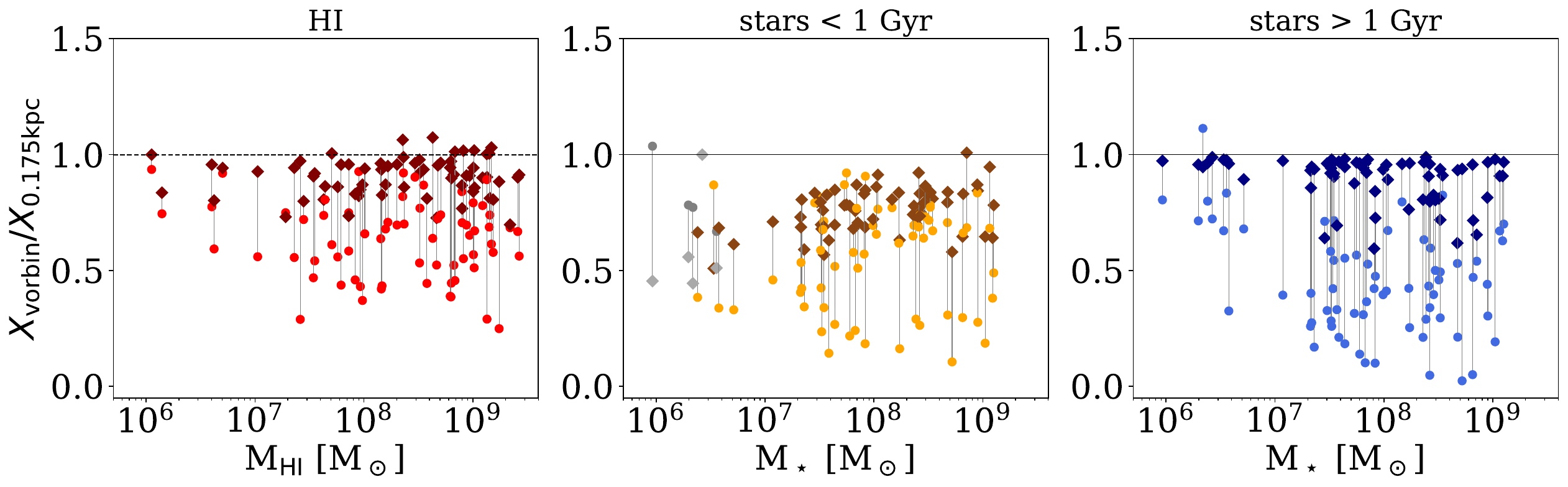}
    \caption{Top panel compares the global rotation velocity and dispersion for the fiducial maps (0.175 kpc in spaxel size) and low-resolution maps (1 kpc in spaxel size). Bottom panel compares the global rotation velocity and dispersion between the fiducial maps and Voronoi maps. Ratios for global rotation velocity are shown with the lighter-shaded circles, and ratios for global velocity dispersion are shown with the darker-shaded diamonds. The grey markers in the middle panel show galaxies with $\leq10$ particles per spaxel in their young stellar maps. As shown in the top row, lower spatial resolution is similar to beam-smearing and results in lower global rotation velocities and higher global dispersions. The bottom row demonstrates how Voronoi binning results in overall lower global rotation velocities \textit{and} lower global dispersions, due to larger spatial bins in the outer regions of the galaxy. Top row: the mean percent difference in V$_\mathrm{global}$ is 30\% for \hi, 43\% for young stars, and 59\% for old stars, while the mean percent difference for $\sigma_\mathrm{global}$ is 21\% for \hi, 48\% for young stars, and 10\% for old stars. Bottom row: the mean percent difference in V$_\mathrm{global}$ is $37\%$ for \hi, $43\%$ for young stars, and $56\%$ for old stars, and the mean percent difference for $\sigma_\mathrm{global}$ is $10\%$ for \hi, $25\%$ for young stars, and $11\%$ for old stars. Younger stars tend to exhibit a larger change due to small number statistics, with less spaxels in their maps compared to \hi ~or old stars.}
    \label{fig:res_mass}
\end{figure*}

In Figure \ref{fig: binsize}, we demonstrate how coarser spatial bins affect the \hi ~kinematic maps for r634-1, which is a disky galaxy (with $\log$(M$_\mathrm{200})=10.72$) at an inclination of $i=52^\circ$. As demonstrated in the top row with V$_\mathrm{LOS}$, larger pixels will `average out' signal within a region and this results in a lower value per pixel. Since we define the global rotation speed as $V_\mathrm{global}= \frac{1}{2}(V_\mathrm{max} - V_\mathrm{min})$, this quantity is sensitive to the outermost regions which have a diluted signal in the map with 1-kpc-sized spaxels. The higher resolution map (left panel) has an inclination-corrected global rotation speed of 97.74 km s$^{-1}$, while the lower resolution (middle) map has an inclination-corrected global rotation speed of 76.77 km s$^{-1}$. With Voronoi binning, larger spatial bins in the outer regions of the galaxy results in an even lower global value of 55.54 km s$^{-1}$.

Similarly, velocity dispersion can be sensitive to spatial resolution. Coarser spatial bins produces a similar effect as beam smearing, in which lower spatial resolution yields a velocity gradient that is less steep, and thus the inferred velocity dispersion can be artificially higher \citep[also discussed in][ for radio and optical observations, respectively]{2015barolo,2009Law}. In the bottom row of Figure \ref{fig: binsize}, the global velocity dispersion increases with lower spatial resolution. The fiducial second moment map of r634-1 exhibits a global LOS dispersion of 10.12 km s$^{-1}$, while the coarser uniform grid has a global value of 11.35 km s$^{-1}$. Velocity dispersion is typically lower in the outer regions of the galaxy, which are weighted more with Voronoi binning compared to a uniform square grid. As shown in Figure \ref{fig: binsize}, the \hi ~Voronoi map for r634-1 has a lower global dispersion of 9.55 km s$^{-1}$.

In Figure \ref{fig:res_mass}, we show how the global LOS velocity and dispersion varies with spatial resolution as a function of galaxy mass. In the top panel, we compare the global quantity of the low-resolution map (1 kpc spaxel) divided by the global quantity of the high-resolution map (0.175 kpc spaxel). We also examine this effect for different tracers: \hi ~(left column), young stars (middle column), and old stars (right column). The ratio of rotation speeds is shown with the lighter-shaded circles, while the ratio of dispersions is shown with darker-shaded diamonds. For the low-resolution maps, rotation velocity can be significantly underestimated. This spatial resolution effect is worse for lower-mass galaxies as the number of resolution elements per galaxy decreases in this test. However, with Voronoi maps, this binning effect is more significant for higher-mass galaxies since larger bins in outer regions `average out' the signal even more. Velocity dispersion is mostly overestimated in the low-resolution maps and Voronoi maps, though the effect of spatial binning is not as significant for dispersion compared to rotation velocity. 

\begin{figure}
    \centering
    \includegraphics[width=\linewidth]{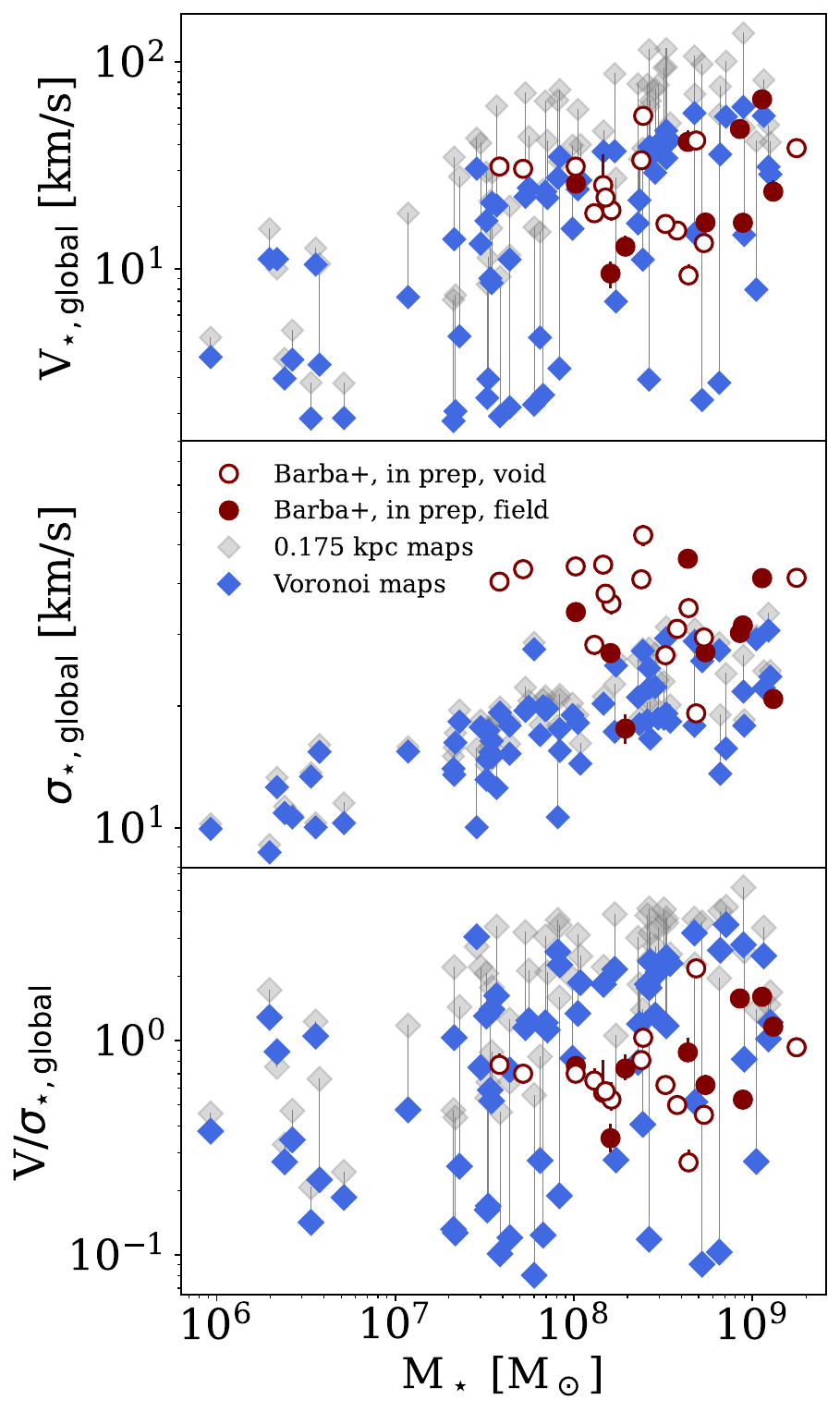}
    \caption{We compare the rotation speed, velocity dispersion, and $V/\sigma_\mathrm{global}$ for old stars when using our fiducial maps (with 0.175 kpc spaxels, grey diamonds) versus maps with Voronoi binning (blue diamonds). We also compare to observational data from \barba, who use Voronoi binning for IFU data. It seems that spatial binning can largely reconcile differences in the measured rotation speed, thus resulting in lower $V/\sigma_\mathrm{global}$.}
    \label{fig:Voronoi_V}
\end{figure}

In Figure \ref{fig:Voronoi_V}, we show how $V_\mathrm{global}$,  $\sigma_\mathrm{global}$, and  $V/\sigma_\mathrm{global}$ changes when using Voronoi binning for old stars in the simulated galaxies. We also plot the rotation speeds from \barba, which are determined from IFU data of stars. This test demonstrates how discrepancies between simulations and observations in $V/\sigma_\mathrm{global}$ may be largely due to spatial resolution and lowered $V_\mathrm{global}$ from Voronoi binning. Interestingly, Voronoi binning seems to slightly decrease the inferred velocity dispersion, whereas Figure \ref{fig:stars_Vsigma} shows some observed dispersions that are \textit{higher} than the simulated dispersions. Therefore, discrepancies between simulated and observed stellar $\sigma_\mathrm{global}$ are likely due to the reasons discussed in Section \ref{stellar_dispersions}, rather than Voronoi binning.

\section{\hi ~Profiles as a Proxy for $V/\sigma$}
\label{HIprofiles}

Current and upcoming surveys will provide \hi ~emission profiles for a large sample of galaxies, and the shape of a given \hi ~profile can provide an indication of whether the galaxy is more dispersion or rotation supported. Double-horned \hi ~profiles are the characteristic shape of a more rotation-supported galaxy, as the gas is moving both away and towards us along the line of sight to produce this distinct shape. Gaussian \hi ~profiles are typically associated with a galaxy that is more dispersion-supported, and this shape is often found in dwarf galaxies. 

\begin{figure*}
    \centering
    \includegraphics[width=\textwidth]{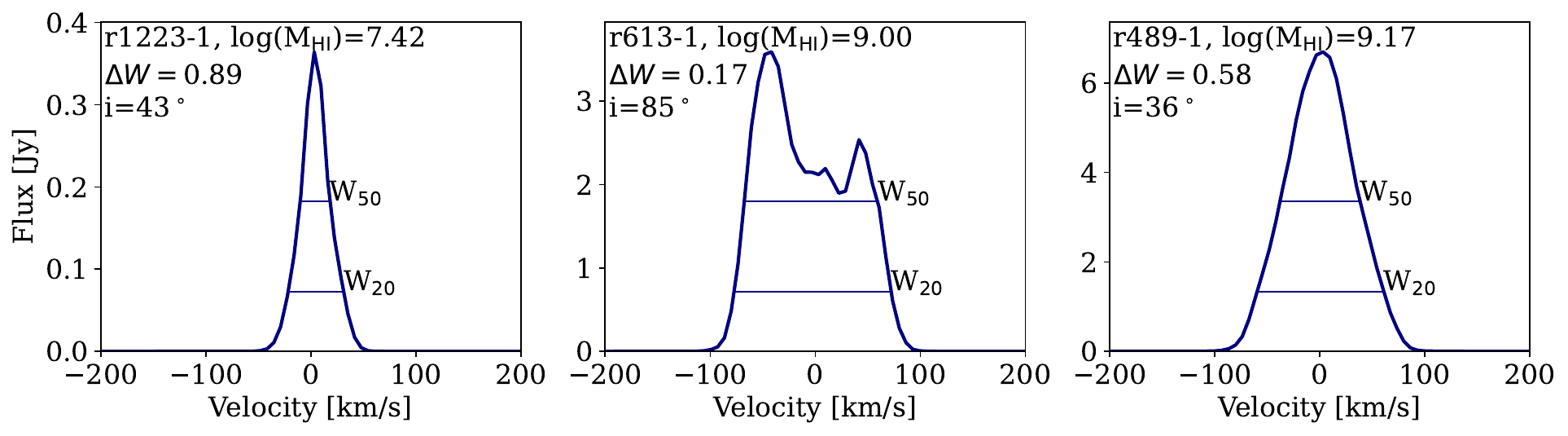}
    \vspace{-6mm}
    \caption{\hi ~profiles for three galaxies in our sample. Left panel shows r1223-1, a compact galaxy (effective radius $<$ 2 kpc). Middle panel shows r613-1, a disky galaxy. Right panel shows r489-1, an irregular galaxy. These \hi ~profiles are each generated at a randomly-drawn inclination angle, shown in Figure \ref{fig:sampleinfo}. We quantify the shape of each \hi ~profile with $\Delta W = (W_\mathrm{20}-W_\mathrm{50})/W_\mathrm{50}$, where $W_\mathrm{20}$ and $W_\mathrm{50}$ are the linewidths at 20\% and 50\% of the peak flux, respectively. Double-horned profiles like that of r613-1 have lower $\Delta W$ values, as the stronger wings result in a lower difference between $W_\mathrm{20}$ and $W_\mathrm{50}$. Gaussian profiles like that of r1223-1 have higher $\Delta W$, as the stronger central peak results in a larger difference between $W_\mathrm{20}$ and $W_\mathrm{50}$. For a galaxy with a strong peak and strong wings, like that of r489-1, the $\Delta W$ value is between the limits set by Gaussian and double-horned profiles.}
    \label{fig:HIprofs}
\end{figure*}

In Figure \ref{fig:HIprofs}, we show the \hi ~emission spectra for three galaxies in our sample, each at a randomly-drawn inclination angle. We have generated idealized \hi ~emission profiles for our simulated dwarf galaxies using the \verb|martini| python code \citep{2024Oman}, which accounts for thermal broadening. We use a channel width of 6.4 km s$^{-1}$ \citep[similar to FASHI,][]{2024Zhang}, with more idealized conditions such as Gaussian noise with an RMS of $3\times10^{-8}$ Jy beam$^{-1}$, and a Gaussian beam that is 12 arcsec in size. For the mock \hi ~profiles, we place each galaxy at a distance of 3 Mpc. 

\begin{figure*}
\vspace{-5mm}
    \centering
    \includegraphics[width=0.49\textwidth]{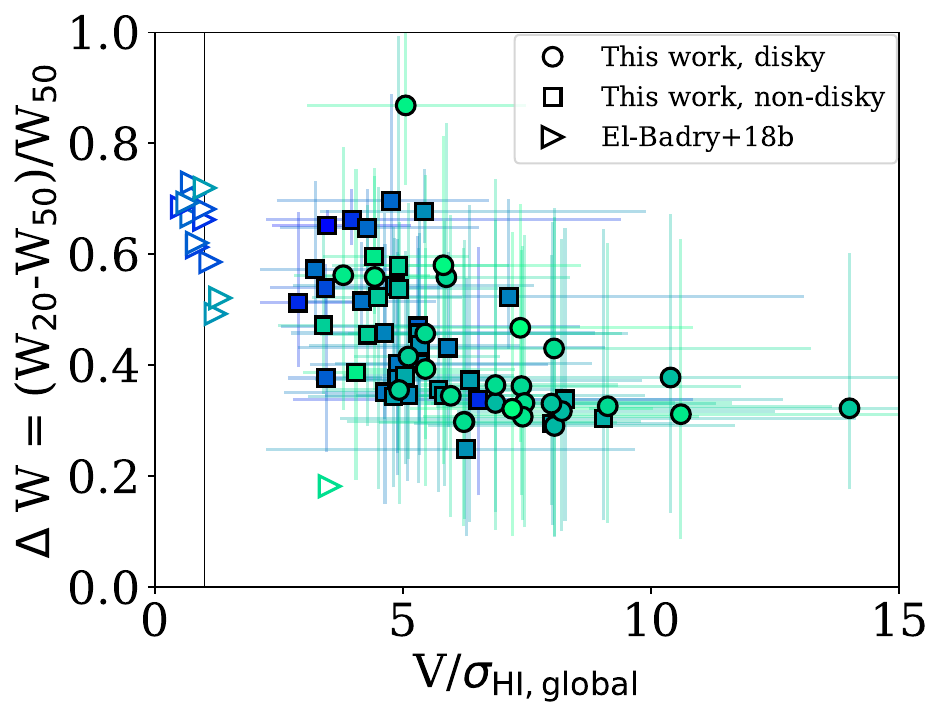} 
    \includegraphics[width=0.503\textwidth]{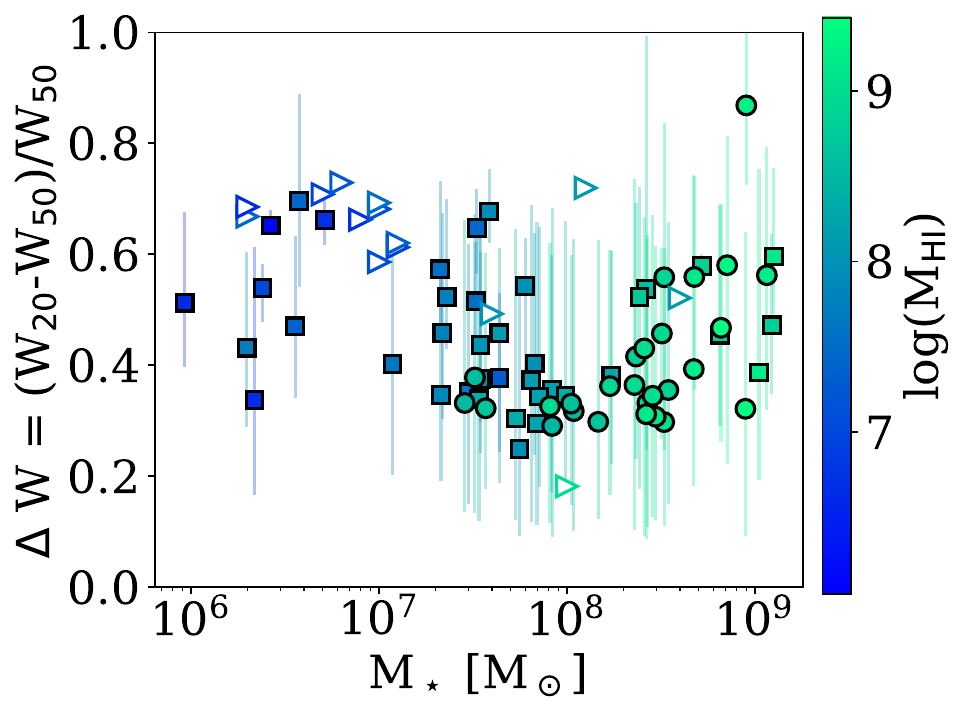} 
    \vspace{-6mm} 
    \caption{Each point is color-coded by \hi ~mass. We use circle markers to show disky galaxies (defined by median ${j}_{z,\star}/j_\mathrm{tot} > 0.5$), and square markers to show non-disky galaxies in our simulated sample. The left panel demonstrates how $\Delta W$ can be a proxy for $V/\sigma_\mathrm{HI,global}$. These x-axis values and error bars slightly differ from those plotted in Figure \ref{fig:HI_Vsigma}. Here we plot the average $V/\sigma_\mathrm{HI,global}$ over all viewing angles, with the horizontal errors extending to the minimum and maximum value over all viewing angles. We plot the average $\Delta W$ over three viewing angles (randomly-drawn inclination, $i=0^\circ$, and $i=90^\circ$), and the vertical error bars extend to the minimum and maximum values over these three viewing angles. Galaxies with $V/\sigma_\mathrm{HI,global}\gtrsim 5$ are more rotation-supported and exhibit \hi ~profiles with $\Delta W\lesssim 0.4$. This relation between $\Delta W$ and $V/\sigma$ is potentially useful to understand the level of rotation support for galaxies which only have spatially unresolved \hi ~data ($\Delta W$). Similar to \citet{2018ElBadryHIprofiles} (triangles), we find an anti-correlation between $\Delta W$ and $V/\sigma_\mathrm{HI,global}$, but in this stellar mass range they mostly found $V/\sigma_\mathrm{HI,global} \lesssim 2$ for their sample, whereas we find $V/\sigma_\mathrm{HI,global}\approx 2.9-14$. The right panel shows $\Delta W$ as a function of galaxy mass, where the data are also color-coded by \hi ~mass. The \hi ~profile shape (quantified by $\Delta W$) is not strictly monotonic with galaxy mass, and can exhibit more diversity ($\Delta W = 0.25-0.87$ for our entire sample). This is likely due to a combination of inclination angle and galaxy morphology, in which face-on viewing angles can produce a more Gaussian \hi ~profile, or irregular galaxies do not exhibit double-horned profiles but instead may have a strong central peak and strong wings (such as r489-1 in Figure \ref{fig:HIprofs}). }
    \label{fig:steepness}
\end{figure*} 

As seen in Figure \ref{fig:HIprofs}, lower-mass galaxies like r1223-1 tend to have a more Gaussian \hi ~profile, while higher-mass galaxies like r613-1 exhibit a double-horned profile. However, not all higher-mass galaxies exhibit a double-horned profile, such as r489-1. We can use the relative linewidth difference $\Delta W = (W_\mathrm{20}-W_\mathrm{50})/W_\mathrm{50}$ to quantify the shape of the \hi ~profile, where $W_\mathrm{20}$ and $W_\mathrm{50}$ are the linewidths at 20\% and 50\% of the peak \hi ~flux. r613-1 has a double-horned profile and a lower value of $\Delta W$ due to its weaker wings and hence more similar values of $W_\mathrm{20}$ and $W_\mathrm{50}$. r1223-1 has a Gaussian \hi ~profile with most of its gas in the central region moving at lower velocities, and therefore has a higher value of $\Delta W$. Most of the gas in r489-1 is also located in its central region, but r489-1 is more extended in size than r1223-1. A more extended \hi ~disk suggests higher velocities, and hence r489-1 exhibits a wider Gaussian \hi ~profile, and hence an intermediate value of $\Delta W$.

 In the left panel of Figure \ref{fig:steepness}, we examine how $\Delta W$ can be a proxy for $V/\sigma$ with the Massive Dwarfs. Each data point is color-coded by the galaxy's total \hi ~mass, and disky galaxies (with median ${j}_{z,\star}/j_\mathrm{tot} > 0.5$) are shown with circle markers while non-disky galaxies are shown with square markers. The x-axis values slightly differ from that of Figure \ref{fig:HI_Vsigma}, since we plot the average $V/\sigma_\mathrm{global}$ over all viewing angles (rather than a randomly-drawn inclination), and the horizontal error bars extend to the minimum and maximum values across all viewing angles. Each galaxy's y-axis value uses the average $\Delta W$ from a randomly-drawn inclination, $i=0^\circ$, and $i=90^\circ$, and vertical error bars extend to the minimum and maximum values across these three viewing angles. Overall, there is an anti-correlation between $\Delta W$ and $V/\sigma_\mathrm{HI,global}$. Galaxies which are more rotation-supported with $V/\sigma\gtrsim 5$ will have weaker wings in their \hi ~profiles and $\Delta W \lesssim 0.4$. Galaxies with lower $V/\sigma_\mathrm{HI,global}$, i.e., with more turbulent, non-circular motions, exhibit more scatter in terms of their \hi ~profile shape (with $\Delta W\approx 0.34-0.7$ for $V/\sigma < 5$).

 The left panel of Figure \ref{fig:steepness} also helps in comparing $V/\sigma_\mathrm{HI,global}$ and morphology between simulations that adopt different physical prescriptions. We compare our trend in $\Delta W$ versus $V/\sigma_\mathrm{HI,global}$ to the result from \citet{2018ElBadryHIprofiles} (their Figure 2) using the FIRE simulations, shown in triangles and color-coded by \hi ~mass \citep[from Table 1 of][]{2018ElBadry}. The FIRE galaxies also exhibit an anti-correlation between $\Delta W$ and $V/\sigma_\mathrm{HI,global}$ \citep{2018ElBadryHIprofiles}. Compared to their results, we find similar ranges in $\Delta W$, but different ranges in $V/\sigma_\mathrm{HI,global}$. The Massive Dwarfs exhibit $V/\sigma_\mathrm{HI,global}=2.9-14$, while \citet{2018ElBadryHIprofiles} found $V/\sigma_\mathrm{HI,global}\lesssim 5$, with all but one of their galaxies $\lesssim 2$. Based on the trend in $\Delta W$ versus $V/\sigma_\mathrm{HI,global}$, it seems that the Massive Dwarfs are more rotation-supported than the galaxies from \citet{2018ElBadryHIprofiles}. We discuss how baryonic feedback models may cause discrepancies between the Massive Dwarfs and FIRE simulations in Section \ref{morphologies}. 

 In the right panel of Figure \ref{fig:steepness}, we find that $\Delta W$ does not have a strictly monotonic relationship with galaxy mass (perhaps unsurprising given the \hi ~profile of r489-1 in Figure \ref{fig:HIprofs}). \citet{2018ElBadryHIprofiles} also noted that the relation between $\Delta W$ and $V/\sigma$ is not simply a result of mass-dependence, given their outlier \verb|m11b| (with M$_\star=10^8$ \msol) exhibiting strong rotation. The Massive Dwarfs simulations exhibit a range of morphologies at a given stellar mass $\gtrsim 10^7$ \msol, both disky and non-disky ~\citep[see also][]{Keith2025,Geda2025}. This mix of morphologies leads to a range of \hi ~profile shapes ($\Delta W = 0.25-0.87$ for our entire sample). Higher $\Delta W$ at higher galaxy masses may be due to inclination angle or irregular morphologies. Double-horned \hi ~profiles can appear more Gaussian at lower inclination angles, or galaxies with irregular morphologies (such as for r489-1) exhibit \hi ~profiles with strong central peaks and strong wings. The \hi ~profile shape likely depends on other factors beyond galaxy mass such as star formation or merger history. In a future work we will examine the diversity in \hi ~profile shape in more detail.

More dwarf galaxies will be discovered with upcoming surveys, in which the \hi ~profiles and linewidths ($W_\mathrm{20}$, $W_\mathrm{50}$) may be the only available information on a galaxy's gas kinematics. Although there is scatter in the relation between $\Delta W$ and $V/\sigma_\mathrm{HI,global}$, the shape of \hi ~profiles may provide a proxy to understand the level of rotation support for a galaxy from spatially unresolved \hi ~data.

\section{Discussion}\label{discussion}

We demonstrate that for a given dwarf galaxy, $V/\sigma$ measured by \hi ~is systematically higher than that of young stars or old stars, and we compare our results to other simulations and observations. Here, we discuss physical reasons beyond spatial binning (Section \ref{sec:resolution}) to reconcile any discrepancies. In Section \ref{stellar_dispersions}, we consider how differences in methodology and sample selection may explain higher stellar dispersions for some observed galaxies relative to the simulations. In Section \ref{feedback}, we discuss how different choices in simulation physics can influence velocity dispersion and hence, galaxy rotation and morphology (Section \ref{morphologies}). In Section \ref{sec:scenarios}, we discuss the implications of our results with respect to dwarf galaxy evolution.

\subsection{Stellar Dispersions in Simulations versus Observations}
\label{stellar_dispersions}

We discuss potential sources of the discrepancy between simulated and observed stellar velocity dispersions (in Figure \ref{fig:stars_Vsigma}). In particular, ten out of \Ngalbarba observed galaxies from \barba ~exhibit velocity dispersions which are higher than the constraints of old stars in the simulated dwarf galaxies at a given stellar mass.

The mismatch could be from different stellar populations probed in observations versus simulations. We use a cutoff of 1 Gyr in age for the simulated galaxies to distinguish `old' versus `young' stars, assuming stellar populations from \citet{2008Marigo} and \citet{2010Girardi}. Meanwhile, the spectra in \barba ~are dominated by the CaII H and K lines from older stars (lifetimes $\sim10$ Gyr), though many of their galaxies also exhibit strong hydrogen Balmer lines (indicative of recent star formation, $<1$~Gyr). If the CaII H and K lines are dominant such that the stellar population is on average older than 1 Gyr in the observed galaxies, this could result in higher velocity dispersions relative to the simulated values.

Differences in sample selection may also influence velocity dispersion. We select galaxies with M$_{\star}=10^6-10^9$ M$_{\odot}$ and M$_{\rm{HI}}>10^6$ M$_{\odot}$. The simulated sample overlaps with that of \barba, which has M$_\star=10^7-10^9$ M$_{\odot}$. However, DIVE galaxies were selected to have high [\oiii]$\lambda4363$ fluxes. Higher [\oiii]$\lambda4363$ fluxes likely mean the DIVE Survey galaxy selection is biased toward low-metallicity systems. As discussed in Section 4.3.1 of \citet{2023delosReyes}, gas-phase metallicities are inversely correlated with specific star formation rate \citep[e.g.,][]{Maiolino2019}, and star formation rate is correlated with velocity dispersion \citep[e.g.,][]{2019Yu,2025Luo}. Therefore, the selection of galaxies with bright [\oiii]$\lambda4363$ fluxes could lead to an observed galaxy sample with higher dispersions than the simulated sample.

Differences in stellar velocity dispersions between observations and the simulated results may be explained through the effect of selection functions and stellar populations, or other systematic uncertainties in the data that are not yet understood. Choices in simulation physics can also influence velocity dispersion. In the next section, we elaborate on what aspects of stellar feedback physics could affect velocity dispersion. 

\subsection{The Impact of Simulation Physics}
\label{feedback}

We discuss how choices related to star formation and feedback in simulations can influence $\sigma$. First, we discuss what influences the initial velocity dispersion of stars at birth, $\sigma_\mathrm{birth}$. Then, we discuss how the choice of supernova feedback model, the choice of subgrid physics, and the implementation of early stellar feedback from young, massive stars may influence $\sigma$.

Previous simulation studies have found that $\sigma_\mathrm{birth}$ depends on the assumed star formation model and resolution of the cold gas. In particular, using better spatial resolution and a higher star formation threshold can result in lower vertical dispersions \citep{2011House,Kumamoto2017} and higher radial dispersions \citep{Kumamoto2017}. By resolving the clumpy, multiphase ISM, including the cold phase, simulations can produce galaxies with relatively thinner disks that exhibit more non-circular motion. As discussed in Section \ref{methods}, the star formation model adopted in the Massive Dwarfs simulation suite requires the presence of H$_\mathrm{2}$, generally ensuring that stars form at densities $n > 100$ cm$^{-3}$. Our star formation physics, in tandem with a force resolution of 87 pc, ensures that stars can only form from dynamically cold gas \citep[see, e.g.,][and discussion therein]{Bird2021}. 

This dispersion floor will influence the dispersion of young stars that we have measured here.  Stars can be dynamically heated quickly after birth (see Appendix \ref{sub:disp_v_time}), so that the dispersions are already higher for a given population even after 1 Gyr.  Hence, the young stellar dispersions we measure here will be influenced by $\sigma_\mathrm{birth}$, but higher than $\sigma_\mathrm{birth}$ due to heating in the first Gyr after formation.  Likewise, $\sigma_\mathrm{birth}$ is not the same as the \hi ~dispersions we have presented here, because stars form from gas with $T < 1000$ K, while \hi ~is dominated by gas closer to $T \sim 10^4$ K.

Thermal injection from supernovae will impact $\sigma$. Models like blastwave supernova feedback \citep{2006Stinson} delay cooling in surrounding gas particles by numerically `turning off' radiative cooling until the end of the snowplow phase. The Massive Dwarfs use superbubble feedback, which accounts for energy input from clustered supernovae and thermal evaporation of the ISM from the cold phase to the hot phase \citep{2014Keller}. Superbubble feedback can have higher mass loading in its outflows, stronger star formation regulation, and more retention of high-angular-momentum gas than blastwave feedback \citep{2014Keller}. In a comparison of two simulated dwarf galaxies with superbubble and blastwave feedback, \citet{2021Mina} found that superbubble feedback resulted in comparable $V/\sigma$ values for stars, and higher $V/\sigma$ values for \hi. Specifically, they found that the superbubble runs had higher \hi ~rotation speeds, and lower \hi ~velocity dispersions. Overall, the superbubble feedback model seems largely responsible for retaining high-angular momentum material in \hi, and therefore higher $V/\sigma$.

SNe may be a dominant source of turbulence at low redshifts \citep[e.g.,][]{Krumholz2018}, and the choice of momentum injection may also influence $\sigma$. Superbubble feedback considers the energy from supernovae to be only in a thermal component, and this thermal energy is transformed into kinetic energy (momentum) through the N-body + SPH code.  Other simulations include momentum injection from supernovae, in addition to thermal energy, particularly if the radius impacted by supernova feedback is unresolved \citep[e.g.,][]{2015Kim}.

In addition to supernovae, a number of simulations implement subgrid physics to capture stellar feedback from young, massive stars with local photoionization, momentum injection from stellar winds, and radiation pressure \citep[e.g.,][]{Wise2012, Aumer2013, Stinson2013, 2014Hopkins, Ceverino2014, Agertz2015, 2018Hopkins}. The Massive Dwarfs currently exclude this feedback, and include only feedback from supernovae. These subgrid feedback implementations can add together in a non-linear way such that it is non-trivial to determine how much energy is coupled to the ISM, and not all simulations will yield similar results, particularly in terms of velocity dispersion \citep[e.g.,][]{Roskar2014, Agertz2015}. However, some have found that inclusion of young stellar feedback could lower the central gas velocity dispersions, at least in Milky Way-sized disks \citep{Agertz2013}. In dwarfs, early stellar feedback can also influence the burstiness of star formation when a stochastic sampling of the initial mass function is adopted \citep{Applebaum2020, Smith2021}. More recently, some simulators have been able to follow radiative hydrodynamics in the formation of cosmological dwarf galaxies \citep[e.g.,][]{Agertz2020, Baumschlager2025, Rey2025}.  In general, these studies have found that the inclusion of radiative transfer leads to less bursty star formation in dwarf galaxies, and \citet{Rey2025} found that it leads to more \hi ~rich dwarfs.  However, the implications for $\sigma$ have yet to be explored in detail. 

Finally, we note that all of these different physics implementation choices can lead to different impacts on potential tracers of $\sigma$.  For example, because of the subgrid model for early stellar feedback used in FIRE, \citet{2025Luo} recently used the ionized gas near young star particles to directly compare dispersions to observed H\,II regions in dwarf galaxies.  Such a comparison is not possible in simulations that lack early stellar feedback.  For example, \citet{Kassin2014} attempted to study the evolution of gas velocity dispersions in Milky Way-mass simulations that did not include early feedback.  They were limited to using cold, potentially star forming gas, or dense gas that had been heated by supernovae.  Neither were a good match to the ionized gas in H\,II regions of galaxies.  Since the Massive Dwarfs do not include early stellar feedback, we do not consider velocity dispersions from ionized gas in this work. Velocity dispersions measured in observations from the emission lines of ionized gas in dwarf galaxies \citep[e.g.,][Barba et al., in prep]{2024Xu,2025Luo} can be up to $\sim 50$ km s$^{-1}$, which is higher than our values for \hi, young stars, and old stars in this work. We also note that our definition of young stars (ages $<$ 1 Gyr) includes stars that would not be traced by emission lines (such as H$\alpha$, which traces stars $\lesssim 10$ Myr in age). Given our results, we expect different velocity dispersions based on the component used; therefore, it makes sense that other studies using ionized gas dispersions may measure higher values than what we measure for \hi ~gas or stars.

\subsection{Comparing Morphologies in Simulations}\label{morphologies}

Given the choices in simulation physics discussed above, there are many reasons why simulations might disagree on $\sigma$. Pinpointing the source of difference in $\sigma$ between simulations is likely to require detailed tests that isolate various aspects of feedback implementations, as well as resolution. We showed in Figure \ref{fig:disky} that $V/\sigma_\mathrm{global}$ is tied to morphology.  In this section, we discuss the effect of feedback choices in terms of different $V/\sigma$ and morphologies between simulations.

We use \hi ~kinematics as a common basis to compare between simulations. In our mass range of dwarf galaxies, we find $\sigma_\mathrm{HI,global}\approx3-20$ km s$^{-1}$ (Figure \ref{fig:HI_Vsigma}), which is consistent with the range from \citet{2019Dutton} ($5-20$ km s$^{-1}$) using the NIHAO simulations \citep{2015Wang}. Other work using the APOSTLE simulations \citep{Oman2019} and the FIREbox simulations \citep{Benavides2025} generally find \hi ~dispersions around $10-20$ km s$^{-1}$. Differences in velocity dispersion can manifest in the overall ratio $V/\sigma$. For galaxies with M$_\mathrm{bary}=10^{8-9}$ M$_\odot$, we find $V/\sigma_\mathrm{HI,global}\sim 4-10$. Our range is consistent with that of NIHAO galaxies \citep[around $1-10$ for V$_\mathrm{max}\sim50-100$ km s$^{-1}$, Figure 11 of][]{2019Dutton} and APOSTLE galaxies \citep[around $4-7$ for M$_\mathrm{bary}\sim10^9$ M$_\odot$][]{Oman2019}. These values are in contrast with the FIRE \citep{2018ElBadryHIprofiles} and FIREbox \citep{Benavides2025} galaxies which exhibit $V/\sigma_\mathrm{HI,global}\lesssim2$ (except for one galaxy in FIRE, \verb|m11b|, which has $V/\sigma_\mathrm{HI,global}\sim4$). The Massive Dwarfs exhibit global \hi ~dispersions and $V/\sigma$ values that are more similar to that of the NIHAO and APOSTLE simulations rather than the FIRE simulations. However, we note that there are other known differences between the Massive Dwarfs, NIHAO, and APOSTLE due to star formation and feedback choices \citep[e.g., NIHAO sizes are larger and APOSTLE simulations do not form dark matter cores;][]{Cruz2025}.

Simulation physics impact $V/\sigma$, and similarly the formation of disks. We demonstrate in Figures  \ref{fig:disky} and \ref{fig:steepness} that the Massive Dwarfs exhibit disky galaxies down to M$_\star\sim3\times10^7$ M$_\odot$ \citep[consistent with][from the same simulation suite]{Keith2025,Geda2025}, and that these disky galaxies exhibit systematically higher $V/\sigma_\mathrm{global}$ than non-disky galaxies at a given mass. Feedback that influences radial motion may influence the formation of disks. \citet{2016ElBadry} found that stars in FIRE could form in gas with strong radial flows, both inward and outward. The feedback driving these flows can drive `breathing modes' in the FIRE simulations \citep{2016ElBadry}, in which the size of the galaxy oscillates with time, and young stars can migrate $\sim$1 kpc within 100 Myr. This feedback either removes angular momentum-rich gas in dwarfs or prevents its accretion \citep{Pandya2020}, and thus prevents the formation of disks \citep{2018ElBadry}. As previously mentioned, \citet{Benavides2025} used the FIREBox simulations \citep{FIREBox} and did not find disks for galaxies with M$_\star < 10^9$ M$_\odot$. In contrast, the Massive Dwarfs and other simulations with similar physics \citep[such as the Marvel/DC Justice League dwarf galaxies][]{Munshi2021} do not exhibit signs of breathing modes. For instance, \citet{Riggs2024} found net radial velocities closer to 0 for young stars (see their Figure 3), and \citet{Geda2025} found no evidence for large size and specific SFR fluctuations. As mentioned in Section \ref{feedback}, superbubble feedback helps retain high angular momentum gas in our simulated dwarf galaxies, and hence higher $V/\sigma$ and the formation of disks.

\subsection{Dwarf Galaxy Scenarios}\label{sec:scenarios}

The fact that our dwarfs are diskier and more rotationally supported than some other simulated and observed dwarfs has implications for the theoretical understanding of whether disky field dwarfs can transform into dispersion-dominated dwarf spheroidal satellites.  As discussed in the Introduction (Section \ref{intro}), initial studies that proposed a transition assumed that infalling dwarfs start with $V/\sigma_\mathrm{global}$ in the range $2-5$.  This idea was called into question when both cosmologically simulated dwarfs \citep{2017Wheeler, Frings2017} and observed field dwarfs \citep{2017Wheeler, 2023delosReyes} demonstrated $V/\sigma_{\star,\mathrm{global}} < 2$.  However, our analysis found several potential biases that could lead to lower $V/\sigma_{\star,\mathrm{global}}$ in observations.  

Our work demonstrates that different baryonic tracers bias the inferred $V/\sigma_\mathrm{global}$. Old stars are always the most dispersion-dominated tracer, as highlighted in Figure \ref{fig:Vsigma_all}.  \citet{2017Wheeler} examined low-mass galaxies in the Local Group using stellar velocity dispersions exclusively of red giant branch and horizontal branch stars, i.e., stars that would fall into our `stars $>$ 1 Gyr' category.  Likewise, \citet{2023delosReyes} and \barba ~measured stellar kinematics from continuum galaxy spectra which are also dominated by older stars.  Understanding how $V/\sigma_\mathrm{global}$ depends on environment using other baryonic tracers could help inform whether dwarf galaxies are indeed born puffy. The relative difference in $V/\sigma_\mathrm{global}$ between baryonic tracers may also be informative of the timescales needed to undergo transformation via tidal stirring, if any.

We found that $V/\sigma_\mathrm{global}$ depends on galaxy mass, and this has implications for interpreting previous studies.  \citet{2017Wheeler} exclusively examined dwarfs with M$_{\star} < 10^7$ M$_{\odot}$.  In that mass range, we also find $V/\sigma_{\star,\mathrm{global}} < 2$, and generally $V/\sigma_{\star,\mathrm{global}} < 1$, meaning that our results are consistent with \citet{2017Wheeler} at these low masses.  Meanwhile, \citet{2023delosReyes} and \barba ~examined field dwarfs up to M$_{\star} \sim 10^9$ M$_{\odot}$. A majority of the observed galaxies in this mass range (M$_{\star} \gtrsim 10^8$ M$_{\odot}$) are dispersion-dominated with $V/\sigma_{\star,\mathrm{global}} < 1$. This is seemingly at odds with the old stars in simulated galaxies that are mostly rotation-dominated. However, this is resolved if we take into account biases due to the resolution imposed on observational studies.

In Section \ref{sec:resolution}, we found that spatial resolution also plays a major role in measuring $V/\sigma_\mathrm{global}$. Lower spaxel resolution (i.e., larger spaxel size), especially in the outer regions of a galaxy, will artificially lower the $V_\mathrm{global}$ measurement. Voronoi binning, which is used to measure stellar dispersions, will also result in lower $\sigma_\mathrm{global}$. The systematic bias due to Voronoi binning is larger for rotation speed than for velocity dispersion. 
The bottom panel of Figure \ref{fig:res_mass} suggests that, for old stars, rotation speeds can be significantly underestimated, particularly for the higher-mass dwarfs, by $\sim50-90$\%.  This may explain why $V_\mathrm{global}$ is as low as 10 km s$^{-1}$ for some of the higher-mass observed dwarfs in Figure \ref{fig:stars_Vsigma}.  

Overall, our results suggest that $V/\sigma_\mathrm{global}$ is likely to be underestimated by the current observations due to bias from spatial binning in IFU data and the spectra being dominated by stellar populations older than 1 Gyr. The effects due to spatial binning and older stellar populations must be further characterized when considering whether dwarf galaxies indeed form as dispersion-dominated objects.

\section{Conclusions} \label{conclusions}
Velocity dispersion and the ratio $V/\sigma_\mathrm{global}$ provide metrics to understand the morphology of a galaxy and its level of rotation versus pressure support. Given the observational difficulty in measuring $V/\sigma_\mathrm{global}$, this parameter has yet to be fully constrained for dwarf galaxies. The Marvelous Massive Dwarfs simulations can reproduce disky dwarf galaxies and rotation curve diversity \citep{Cruz2025, Geda2025,Keith2025} and the \hi ~content of dwarfs \citep{2025Ruan}, and with these successes we examine $V/\sigma_\mathrm{global}$ as a function of mass. We calculate $V/\sigma_\mathrm{global}$ for \Ngal galaxies with M$_\star=10^6-10^9$ \msol ~from the Massive Dwarfs and compare mass trends for different baryonic components. We produce line-of-sight maps for rotation speed and velocity dispersion and determine global quantities over 31 viewing angles. We compare rotation speed, velocity dispersion, and $V/\sigma_\mathrm{global}$ as a function of galaxy mass for our simulated galaxies and constraints from observations. The results for \hi ~kinematics are shown in Figure \ref{fig:HI_Vsigma} and the results for stellar kinematics are shown in Figure \ref{fig:stars_Vsigma}. Our key results are:
\begin{itemize}
    \item We find that $V/\sigma$ increases with galaxy mass (Figure \ref{fig:Vsigma_all}). \hi ~gas and young stars (ages $<$ 1 Gyr) are more rotation supported in this mass regime, with $V/\sigma_\mathrm{global}\approx 1-13$. Old stars (ages $> 1$ Gyr) are more dispersion supported with $V/\sigma_\mathrm{global}\approx 0.2-5$. This is consistent with the idea that stars become more dispersion supported over time due to dynamical heating. In summary, the level of rotation support measured for a given galaxy depends on its mass and which baryonic component is being used to trace the kinematics. 

    \item In terms of the stellar kinematics, some observed galaxies exhibit higher dispersions and lower rotation speeds than our simulated sample. Higher stellar dispersions may be due to differences in the selection functions and stellar populations (Section \ref{stellar_dispersions}). Lower rotation speeds in observations are likely due to lower spatial resolution or Voronoi binning (Section \ref{sec:resolution}, Figure \ref{fig:res_mass}). The effect on rotation speed due to Voronoi binning has more of an impact to lower $V/\sigma_\mathrm{global}$ than dispersion (Figure \ref{fig:Voronoi_V}).

    \item We find that disky and oblate galaxies exhibit systematically higher $V/\sigma_\mathrm{global}$ than non-disky galaxies at a given baryonic mass (Figure \ref{fig:disky}). The relative difference in $V/\sigma$ between disky and non-disky galaxies is more apparent in \hi ~gas and young stars.
    
    \item Upcoming surveys will measure \hi ~emission profiles for many more galaxies, offering the potential opportunity to quantify rotation support for large samples. The \hi ~profile shape can be quantified through the relative linewidth difference $\Delta W = (W_\mathrm{20}-W_\mathrm{50})/W_\mathrm{50}$, and this metric serves as a proxy for $V/\sigma_\mathrm{global}$. Similar to \citet{2018ElBadryHIprofiles}, we find that $\Delta W$ and $V/\sigma_\mathrm{HI,global}$ are inversely proportional, but we find higher $V/\sigma_\mathrm{HI,global}$ values (up to 14, Figure \ref{fig:steepness}), and a wider spread in $\Delta W$. 
    
    \item Related, diverse morphologies for the simulated dwarf galaxies manifest in terms of scatter in $\Delta W$ (right panel of Figure \ref{fig:steepness}). At a fixed galaxy mass, morphology and $\Delta W$ can vary widely.  We recover a range of double-horned and Gaussian \hi ~profile shapes with $\Delta W = 0.25-0.87$.
\end{itemize}

Previous observations suggest that dSphs may be born dispersion-dominated due to low $V/\sigma_\mathrm{global}$ and a lack of trend with environment. Our results suggest that $V/\sigma_\mathrm{global}<1$ may be the case for dwarf galaxies and their old stars, but the same galaxy may exhibit strong rotation support in \hi ~gas or young stars. As stars are born from cold gas and undergo dynamical heating over time, understanding $V/\sigma_\mathrm{global}$ as a function of environment \textit{and} for different baryonic tracers will help us understand how dwarf galaxies evolve. 

We have found that the measurements of $V/\sigma_\mathrm{global}$ using older stellar populations, as well as binning of IFU spaxels to achieve sufficient signal-to-noise, might have biased previous observational studies to lower determinations of $V/\sigma_\mathrm{global}$.  We predict that measurements of young stellar populations and \hi ~will yield higher $V/\sigma_\mathrm{global}$ values.  Given our trend of $V/\sigma_\mathrm{global}$ with mass, our results are not in contradiction with previous results that low-mass field dwarfs are born dispersion-dominated.  At M$_{\star} \lesssim 10^7$ M$_{\odot}$, we generally find that $V/\sigma_{\star,\mathrm{global}} < 1$.  

 The Massive Dwarfs successfully produce a mix of morphologies, including disks, down to stellar masses of $10^7$ M$_\odot$. We determine disks are correlated with higher $V/\sigma_\mathrm{global}$. We find higher $V/\sigma_\mathrm{global}$ values than previous simulations and stellar observations. Understanding star formation and feedback through observable signatures like velocity dispersion remain essential to probe galaxy formation, constrain galaxy formation models, and make predictions for the large surveys to come.

\begin{acknowledgments}
DR and AMB thank Andrew Baker for useful discussions about this work. Resources supporting this work were provided by the NASA High-End Computing (HEC) Program through the NASA Advanced Supercomputing (NAS) Division at Ames Research Center. Some of the simulations were performed using resources made available by the Flatiron Institute. The Flatiron Institute is a division of the Simons Foundation. This work used Stampede2 at the Texas Advanced Computing Center (TACC) through allocation MCA94P018 from the Advanced Cyberinfrastructure Coordination Ecosystem: Services \& Support (ACCESS) program, which is supported by U.S. National Science Foundation grants \#2138259, \#2138286, \#2138307, \#2137603, and \#2138296. DR is supported by the NASA FINESST Fellowship (grant 80NSSC25K0300-135900). AMB acknowledges support from NSF grant AST-2510900 and from grant FI-CCA-Research-00011826 from the Simons Foundation. AHGP acknowledges support from U.S. National Science Foundation grant AST-2510899.  BWK acknowledges support provided by NASA through a grant from the Space Telescope Science Institute, through grant HST AR-17547.  JWW acknowledges the support of an NSERC discovery grant. 
\end{acknowledgments}

\software{numpy \citep{numpy}, scipy \citep{2020SciPy-NMeth}, pandas \citep{pandas}, pynbody \citep{2013Pontzen}, martini \citep{2024Oman}}

\appendix
\section{Velocity Dispersion Over Time}
\label{sub:disp_v_time}

\begin{figure}[ht!]
    \centering
    \includegraphics[width = \linewidth]{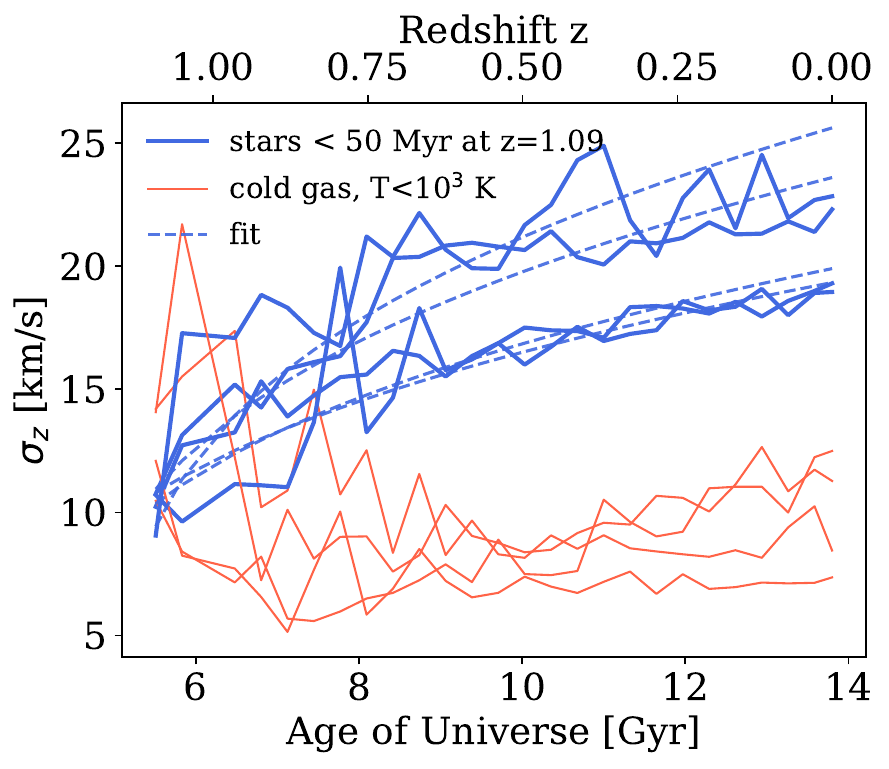}
    \caption{1D out-of-plane velocity dispersion over time (from $z=1.09$ to $z=0$) for stars (blue lines) and cold ISM gas with $T<10^3$ K (red lines). We choose this temperature cut-off to focus on the star-forming gas, since \hi ~is dominated by gas with $T\sim10^4$ K. The dashed lines are fits to Equation \ref{scattering}, which describes velocity dispersion due to midplane scattering with a diffusion coefficient of $\gamma$. As a proof-of-concept, we demonstrate how young stars undergo dynamical heating over time for disky galaxies which have not had recent major mergers. With higher $\sigma$ of old stars, we therefore measure lower $V/\sigma$ values.}
    \label{fig: with_age}
\end{figure}
 
 Here we examine how velocity dispersion varies over time for young stars compared to the ISM they were born from. Dynamical heating can occur through various channels including interactions with satellites \citep{2012Helmi} or spiral arms and molecular clouds \citep{2015Kruijssen}. We focus on the latter and examine the stellar dispersions of galaxies which have not had mergers (with a mass ratio of $\leq$ 1:10) since $z\approx1$. We study this for galaxies r523-1, r571-1, r569-1, and r634-1, which are disky and have $\log($M$_\star)=8.42-8.51$ and $\log($M$_\mathrm{HI})=8.94-9.13$. We trace particles of young stars which have ages $<100$ Myr at $z=1.09$ and measure the velocity dispersion of those same set of stars from then to $z=0$. In particular, we assess how dynamical heating may increase motion out of the plane, so we align the galaxy face-on ($i=0^\circ$) in the x-y plane and measure velocity dispersion in the z-direction ($\sigma_z$). 

In Figure \ref{fig: with_age} we show $\sigma_z$ as a function of time for the selected population of stars (blue). We find that stellar velocity dispersions of these coeval populations indeed increase over cosmic time. We fit these stellar dispersions to an analytic model (blue dashed lines) for scattering of a stellar population on circular orbits in the plane of the disk \citep{1951Spitzer,1977Wielen}. The LOS stellar velocity dispersion $\sigma_z$ from the time that the stars were born ($t$) follows the relation 

\begin{equation}
    \sigma_z = \left( \sigma_\mathrm{ISM}^3 + \frac{3}{2}\gamma t \right)^{1/3}\text{.}
     \label{scattering}
\end{equation}

In this relation, $\sigma_\mathrm{ISM}$ is the velocity dispersion of the ISM gas the stars were born from, and past studies typically assume a floor of $\sigma_\mathrm{ISM}\approx 5-10$ km s$^{-1}$ \citep{1977Wielen}. We set $\sigma_\mathrm{ISM}$ equal to the stellar velocity dispersion at $z= 1.09$, which ranges from $\sim 9-10$ km s$^{-1}$.  $\gamma$ is the velocity-dependent diffusion coefficient of the scattering process, in units of (km s$^{-1}$)$^3$ yr$^{-1}$. For galaxies in the Local Group, the diffusion coefficient has been constrained to be $\gamma = 6\times 10^{-8} - 1\times10^{-5}$ (km s$^{-1}$)$^3$ yr$^{-1}$ \citep{2017Leaman}. Based on the stellar dispersions of our simulated dwarf galaxies shown in Figure \ref{fig: with_age}, we find diffusion coefficients of $\gamma\approx 5\times 10^{-7}-1\times10^{-6}$ (km s$^{-1}$)$^3$ yr$^{-1}$.
 
We also measure $\sigma_z$ for cold gas with $T<10^3$ K and show its $\sigma_z$ as a function of time (red lines). This temperature is slightly cooler than the majority of \hi ~in the ISM, but we choose a cooler cut in order to align more closely with the gas that stars are forming from.  Still, not all of the  cold gas in Figure \ref{fig: with_age} is cool enough to trace the star-forming ISM, sometimes resulting in \hi ~dispersions at $z= 1.09$ that are higher than the dispersions of the young stars.

\vspace{-3mm}
\bibliography{citations}{}
\bibliographystyle{aasjournalv7}
\end{document}